\documentclass[a4paper,12pt]{article}

\pdfoutput=1 % if you are submitting a pdflatex (i.e.\ if you have
\usepackage[hmargin=.71in,vmargin=1.1in]{geometry}
\usepackage{indentfirst}
\usepackage{amsfonts}
\usepackage{mathrsfs}
\usepackage{amsmath}
\usepackage{mathabx}
\usepackage{amssymb}
\usepackage{authblk}
\usepackage{cite}
\usepackage{xcolor}
\usepackage{mathtools}
\usepackage{tensor}
\usepackage{physics}
\usepackage{graphicx}% Include figure files
\usepackage{bm}% bold math
\usepackage{upgreek}
\usepackage{braket}
\usepackage{color,soul}
\usepackage{csquotes}
\usepackage{caption}
\usepackage{subcaption}
\usepackage{slashed}
\usepackage{mathtools}
\usepackage{multirow}%table
\usepackage{tabularx}
\usepackage{adjustbox}

\usepackage[bookmarksnumbered=true,bookmarksopen=true]{hyperref}
 \hypersetup{colorlinks,
             linkcolor=[rgb]{0,0.3,0.6}, 
             citecolor=[rgb]{0,0.3,0.6}, 
             urlcolor=[rgb]{0,0.3,0.6}}

\newcommand\mi{\mathrm{i}}
\newcommand\me{\mathrm{e}}
\newcommand\const{\text{const}}

\newcommand{\dif}{\mathrm{d}}

\newtheorem{remark}{Remark}

\usepackage{tikz}
\usetikzlibrary{matrix}

\begin{document}

\title{\Large\textbf{Comparison of Quasinormal Modes of Black Holes in $f(\mathbb{T})$ and $f(\mathbb{Q})$ Gravity}}

\author[a]{Zhen-Xiao Zhang\thanks{zx.zhang@mail.nankai.edu.cn}}	
\author[b]{Chen Lan\thanks{stlanchen@126.com}}
\author[a]{Yan-Gang Miao\thanks{Corresponding author: miaoyg@nankai.edu.cn}}

\affil[a]{\normalsize{\em School of Physics, Nankai University, 94 Weijin Road, Tianjin 300071, China}}
\affil[b]{\normalsize{\em Department of Physics, Yantai University, 30 Qingquan Road, Yantai 264005, China}}

\date{ }

\maketitle

\begin{abstract}
%\noindent

We investigate the quasinormal modes of static and 
spherically symmetric black holes in vacuum within 
the framework of $f(\mathbb{Q}) = \mathbb{Q} + \alpha \mathbb{Q}^2$ gravity, and compare them
with those in  $f(\mathbb{T}) = \mathbb{T} + \alpha \mathbb{T}^2$  gravity. 
Based on the Symmetric Teleparallel Equivalent of General Relativity, we notice that
the gravitational effects arise from non-metricity
(the covariant derivative of metrics) in $f(\mathbb{Q})$  gravity rather than curvature in $f(R)$ or torsion in $f(\mathbb{T})$.
Using the finite difference method and the sixth-order WKB method, we compute the quasinormal modes of massless scalar field and electromagnetic field perturbations. Tables of quasinormal frequencies for various parameter configurations are provided based on the sixth-order WKB method.
Our findings reveal the differences in the quasinormal modes of black holes in  $f(\mathbb{Q})$  gravity 
compared to those in  $f(R)$  and  $f(\mathbb{T})$  gravity. 
This variation demonstrates the impact of different parameter values, 
offering insights into the characteristics of  $f(\mathbb{Q})$  gravity.
These results provide the theoretical groundwork for assessing alternative gravities' viability through gravitational wave data, 
and aid probably in picking out the alternative gravity theory that best aligns with the empirical reality.

\end{abstract}

\tableofcontents

%%%%%%%%%%%%%%%%%%%%%%%%%%%%%%%%%%%%%%%%%%%%%%%%%%%%%%%%%%%%
\section{Introduction}
%%%%%%%%%%%%%%%%%%%%%%%%%%%%%%%%%%%%%%%%%%%%%%%%%%%%%%%%%%%%

Einstein’s General Relativity (GR)~\cite{Einstein:1915ca} has achieved the remarkable success, 
yet certain fundamental issues remain unresolved, 
such as the accelerated expansion of the universe~\cite{SupernovaSearchTeam:1998fmf,Niedermann:2019olb,Colgain:2021pmf,Nojiri:2021jxf,Landim:2021www}.
The conventional approach to address this issue involves introducing the energy with negative pressure, 
referred to as dark energy, or incorporating a cosmological constant into the action, 
which results in an asymptotically non-flat universe~\cite{Sotiriou:2008rp}. 
However, the lack of direct observational evidences for dark energy leaves this problem unsolved, 
prompting to alternative approaches. Modifying the gravitational theory itself~\cite{Nojiri:2017ncd,Frusciante:2019xia,Oikonomou:2020qah,Solanki:2022ccf}, 
instead of introducing exotic matter fields, emerges as a natural solution.

The idea of modifying gravity could trace back to 1919 when Weyl proposed~\cite{Weyl:1993kh} adding higher-order correction terms to the gravitational action, 
albeit for purely mathematical reasons at that time. 
Over time, the need for modifications became more apparent in order to renormalize GR for its consistency with quantum theories \cite{Utiyama:1962sn,CANTATA:2021asi,Reyes:2022mvm,Shankaranarayanan:2022wbx,Singh:2022jue} 
and to understand the observational deviations in both the early and late universe \cite{Guth:1980zm,SupernovaSearchTeam:1998fmf,Creminelli:2017sry,Braglia:2020auw,Adi:2020qqf}. 
These developments established the groundwork for exploring alternatives to GR, 
such as $f(R)$ gravity \cite{Starobinsky:1980te,Sotiriou:2008rp}, one of the earliest modified gravity theories.

Similar to GR, $f(R)$ gravity is formulated within the framework of Riemannian geometry. The key difference lies in replacing the Ricci scalar $R$ in the gravitational action,
\begin{equation}
S_{\rm GR}=\frac{1}{2\kappa} \int \dif^4 x \sqrt{-g}\; R,\label{GRaction}
\end{equation}
with a general function $f(R)$. This substitution naturally modifies the field equations and leads to new predictions.

The Riemannian geometry itself is a specific case of affine geometry, 
characterized by two core assumptions: the absence of torsion and the vanishing covariant derivative of metrics. 
These conditions are expressed mathematically by
\begin{subequations}
    \begin{equation}
        T^\alpha_{\enspace\mu\nu}=\Gamma^\alpha_{\enspace\mu\nu}-\Gamma^\alpha_{\enspace\nu\mu}=0,
    \end{equation}
    \begin{equation}
        Q_{\alpha\mu\nu}=\nabla_\alpha g_{\mu\nu}=0.
    \end{equation}
\end{subequations}
Relaxing these assumptions gives rise to alternative geometric frameworks, 
such as those underlying $f(\mathbb{Q})$ gravity \cite{BeltranJimenez:2017tkd,Bahamonde:2021gfp,Heisenberg:2023lru}, 
which seek to address unresolved issues in cosmology and gravity \cite{BeltranJimenez:2019tme,Zhao:2022gxl,Dimakis:2024fan}.

By assuming either $R^\alpha_{\ \mu\nu\rho} = 0$ and $Q_{\alpha\mu\nu} = 0$, 
or $R^\alpha_{\ \mu\nu\rho} = 0$ and  $T^\alpha_{\ \mu\nu} = 0$, 
one can derive the Teleparallel Equivalent of General Relativity (TEGR) \cite{Golovnev:2018red} 
and the Symmetric Teleparallel Equivalent of General Relativity (STEGR) \cite{Heisenberg:2023lru}, respectively. 
These frameworks replace the Ricci scalar $R$ in the gravitational action with the scalars $\mathbb{T}$ and $\mathbb{Q}$, respectively. 
Notably, the TEGR and STEGR are mathematically equivalent to GR, yielding identical field equations and conclusions. 
However, their nonlinear generalizations, replacing  $\mathbb{T}$ and $\mathbb{Q}$  with arbitrary functions  $f(\mathbb{T})$  and  $f(\mathbb{Q})$, introduce dynamics that differ fundamentally from GR \cite{Capozziello:2022zzh,Capozziello:2023vne}.
It is important to note that most models exhibit strong coupling in their solutions, as shown in Refs.~\cite{BeltranJimenez:2020fvy,BeltranJimenez:2021auj}. Further analysis reveals a ghost instability in their dynamical degrees of freedom \cite{Gomes:2023tur}. However, viable improvements have been identified within teleparallel gravity, where a class of consistent and stable theories exists \cite{Bello-Morales:2024vqk}. 
Due to the complexity of improved $f(\mathbb{Q})$ gravity, the construction of BH solutions within this framework remains a challenging task. Nevertheless, the original $f(\mathbb{Q})$ gravity can serve as a reasonable approximation to the improved version for investigating spherically symmetric or axisymmetric BHs.  Moreover, it may offer valuable insights into the distinctive features to explore a variety of black hole configurations. This perspective is exemplified by the analysis of spherically symmetric BHs presented in this work.

The motivation for exploring alternatives to GR, such as the $f(\mathbb{T})$ and$f(\mathbb{Q})$ theories investigated in this paper, is twofold: phenomenological and fundamental. Phenomenologically, GR, despite its immense success, requires the introduction of concepts like dark energy and cosmic inflation to align with cosmological observations. Modified gravity theories provide a compelling alternative for these problems \cite{Heisenberg:2023lru}, by suggesting that these phenomena might not arise from new and exotic forms of matter or energy, but from a deviation of gravity itself from GR on large scales. This approach, modifying the law of gravity rather than the inventory of cosmic matter, is considered to be an elegant potential solution. Fundamentally, $f(\mathbb{T})$ gravity and $f(\mathbb{Q})$ gravity are the two most fundamental directions for modifying GR based on a geometric framework. To study them is of fundamental significance for discussing theories that modify GR within a geometric framework.

Black holes (BHs), as the simplest object in the universe \cite{Chandrasekhar:1985kt,Frolov:1998wf}, 
are uniquely suited for testing gravitational theories due to their description by only a few basic parameters \cite{Hamil:2025vey,Karmakar:2023mhs}. 
When perturbed, BHs emit gravitational waves \cite{Cai:2017cbj,Hohmann:2018wxu}, whose evolution goes through three stages \cite{Nollert:1999ji,Berti:2009kk,Konoplya:2011qq}. 
The first stage, which depends on the way of perturbations, is short-lived. 
The second stage, known as the ring-down phase, 
involves damped oscillations as BHs radiate energy. 
The final stage is characterized by a power-law tail, 
where the intensity of gravitational waves diminishes, 
and the spacetime returns to equilibrium. 
The time dependence during the ring-down phase can be expressed as  $\me^{-\mi \omega t}$, where  $\omega$ is the quasinormal mode (QNM) frequency \cite{Maggiore:2018sht}. 
This complex frequency encodes the oscillation frequency (real part) and decay rate (imaginary part). 
Importantly, QNMs depend only on the BH’s intrinsic properties and are independent of the way of perturbations, 
making them highly valuable for observations.

The study of BH QNMs provides a robust tool for testing gravitational theories \cite{Zhu:2014sya,Bhattacharyya:2017tyc,Guo:2020caw,Zhu:2023rrx,Xia:2023zlf,Yang:2024prm,Konoplya:2025hgp}, 
as observational data can be directly compared to theoretical predictions. 
Numerous calculations have been made for the QNMs in $f(R)$ gravity and  $f(\mathbb{T})$  gravity \cite{Aragon:2020xtm,Younesizadeh:2021udk,Ovgun:2018gwt,Datta:2019npq,Karmakar:2024xwr,Zhao:2022gxl}. However, the research on QNMs in  $f(\mathbb{Q})$  gravity remains limited, primarily owing to the complexity of $f(\mathbb{Q})$ BH solutions. 
Previous works by Gogoi \textit{et al.} \cite{Gogoi:2023kjt} and Ahmad \textit{et al.} \cite{Al-Badawi:2024iqv} have explored the QNMs for static and spherically symmetric BHs in  $f(\mathbb{Q})$  gravity, 
assuming a constant scalar  $\mathbb{Q}$  throughout spacetime \cite{Calza:2022mwt}. 
While this assumption simplifies calculations \cite{Calza:2023hhi}, it imposes constraints on the solution space, potentially reducing the range of solutions compared to GR with a cosmological constant \cite{DAmbrosio:2021zpm}. As a result, the solutions may appear observationally indistinguishable from GR in specific configurations.

Furthermore, some other studies often lack direct comparisons of QNMs among different modified gravity theories, 
but such comparisons are crucial for distinguishing these gravity theories observationally. 
To address the shortcoming, we examine at first the static and spherically symmetric vacuum solution in  $f(\mathbb{Q})$  gravity proposed by Fabio \textit{et al.} \cite{DAmbrosio:2021zpm}. 
This solution is based on the functional form  $f(\mathbb{Q}) = \mathbb{Q} + \alpha \mathbb{Q}^2$, where  $\alpha$  is a small parameter, 
and the metric is expressed as a perturbation of the Schwarzschild solution.
We then analyze the QNMs associated with these BH solutions, 
focusing on their differences in comparison with those in $f(\mathbb{T})$  gravity. 
Our study aims to provide theoretical insights and observational strategies for distinguishing these two modified gravity frameworks.

The primary goal of this work is therefore to provide the systematic analysis of QNMs and gravitational waveforms in this $f(\mathbb{Q})$ black hole background, and more importantly, to uncover its unique dynamical signatures. As we will demonstrate, our analysis reveals novel features, such as a ``double-value'' spectral structure of the QNMs, which provides a distinctive observational fingerprint for this theory and offers a clear path to distinguish it from its theoretical counterparts like $f(\mathbb{T})$ gravity.

The structure of the present work is organized as follows. 
Section \ref{sec:f(q)} provides a concise introduction to $f(\mathbb{Q})$ gravity. 
We then present the BH solutions of $f(\mathbb{Q})$ gravity in Section \ref{sec:sssol}, along with an initial analysis and necessary pretreatments. 
Section \ref{sec:numeric} explores the QNMs calculated under varying parameters in terms of the sixth-order WKB method and numerical integration. Section \ref{sec:comparison} compares these results with those obtained from  $f(\mathbb{T})$  gravity \cite{Zhao:2022gxl}. Finally, Section \ref{sec:conclusion} summarizes our findings and conclusions. Throughout this paper, we adopt the natural unit system, setting the gravitational constant $G$ and the speed of light $c$ to be unit.

%%%%%%%%%%%%%%%%%%%%%%%%%%%%%%%%%%%%%%%%%%%%%%%%%%%%%
\section{A brief introduction of \texorpdfstring{$f(\mathbb{Q})$}{f(Q)} gravity}
\label{sec:f(q)}
%%%%%%%%%%%%%%%%%%%%%%%%%%%%%%%%%%%%%%%%%%%%%%%%%%%%%

In this section, we provide a concise overview of  $f(\mathbb{Q})$  gravity 
and establish the notations used throughout the paper.

The framework of  $f(\mathbb{Q})$  gravity is built on metric-affine geometry, 
characterized by the triplet $(\mathcal{M}, g_{\mu\nu}, \Gamma^\alpha_{\ \mu\nu})$, 
where $\mathcal{M}$ is a four-dimensional manifold, $g_{\mu\nu}$ is the metric 
with a signature $(-1, +1, +1, +1)$, and $\Gamma^\alpha_{\ \mu\nu}$ represents the affine connection. 
This geometry allows for three independent geometric quantities: 
curvature $R^\alpha_{\ \beta\mu\nu}$, torsion $T^\alpha_{\ \mu\nu}$, 
and non-metricity $Q_{\alpha\mu\nu}$, defined as
\begin{subequations}
\begin{equation}
R^\alpha_{\ \beta\mu\nu} = \partial_\mu \Gamma^\alpha_{\ \nu\beta} - \partial_\nu \Gamma^\alpha_{\ \mu\beta} + \Gamma^\alpha_{\ \mu\lambda} \Gamma^\lambda_{\ \nu\beta} - \Gamma^\alpha_{\ \nu\lambda} \Gamma^\lambda_{\ \mu\beta},   
\end{equation}
\begin{equation}
T^\alpha_{\ \mu\nu} = \Gamma^\alpha_{\ \mu\nu} - \Gamma^\alpha_{\ \nu\mu},
\end{equation}
\begin{equation}
Q_{\alpha\mu\nu} = \nabla_\alpha g_{\mu\nu} = \partial_\alpha g_{\mu\nu} - \Gamma^\lambda_{\ \alpha\mu} g_{\nu\lambda} - \Gamma^\lambda_{\ \alpha\nu} g_{\mu\lambda}.
\end{equation}
\end{subequations}
In the STEGR, the curvature and torsion tensors vanish, i.e.,
\begin{equation}
R^\alpha_{\ \mu\nu\rho} = 0, \qquad T^\alpha_{\ \mu\nu} = 0,   
\end{equation}
leaving the non-metricity tensor $Q_{\alpha\mu\nu}$ as the sole non-zero geometric object. 
The symmetry of $Q_{\alpha\mu\nu}$ in its last two indices allows for two independent traces,
\begin{equation}
Q_\alpha = Q_{\alpha\ \lambda}^{\ \lambda}, \qquad \overline{Q}_\alpha = Q^\lambda_{\ \alpha\lambda}.
\end{equation}

To construct the action, a suitable scalar $\mathbb{Q}$ is required. 
It is expressed as a linear combination of all five possible scalars contracted from $Q_{\alpha\mu\nu}$,
\begin{equation}
 \mathbb{Q} = b_1 Q_{\alpha\mu\nu}Q^{\alpha\mu\nu} + b_2 Q_{\alpha\mu\nu}Q^{\nu\mu\alpha} + b_3 Q_\mu Q^\mu + b_4 \overline{Q}\mu \overline{Q}^\mu + b_5 Q\mu \overline{Q}^\mu,   
\end{equation}
where $b_i$'s ($i = 1, \ldots, 5$) are constants. 
To ensure the field equations recover GR, the constants in the scalar $\mathbb{Q}$ are uniquely determined, resulting in the expression,
\begin{equation}
\mathbb{Q} = \frac{1}{4} Q_{\alpha\mu\nu} Q^{\alpha\mu\nu} - \frac{1}{2} Q_{\alpha\mu\nu} Q^{\nu\mu\alpha} - \frac{1}{4} Q_\mu Q^\mu + \frac{1}{2} Q_\mu \overline{Q}^\mu.
\end{equation}
This formulation ensures that the action of the STEGR,
\begin{equation}
    S=\frac{1}{2\kappa}\int\dif^4 x \sqrt{-g}\;\mathbb{Q},
\end{equation}
is equivalent to the Hilbert-Einstein action, see Eq.~\eqref{GRaction},
up to a boundary term.
In this context, the Ricci scalar $R$, derived from the Levi-Civita connection, is related to $\mathbb{Q}$ by
\begin{equation}
R = \mathbb{Q} + \mathcal{D}_\alpha (Q^\alpha - \overline{Q}^\alpha),
\end{equation}
where $\mathcal{D}_\alpha$ denotes the covariant derivative. The Levi-Civita connection is given by
\begin{equation}
  \left\{^\alpha_{\enspace\mu\nu}\right\}=\frac{1}{2}g^{\alpha\lambda}\left(\partial_\mu g_{\lambda\nu}+\partial_\nu g_{\lambda\mu}-\partial_\lambda g_{\mu\nu}\right).
\end{equation}
By generalizing the action to $f(\mathbb{Q})$, the gravitational action becomes
\begin{equation}
S = \frac{1}{2\kappa}\int \dif^4 x \sqrt{-g} \,  f(\mathbb{Q}) .
\end{equation}
The metric field equations are derived by varying the action with respect to the metric $g_{\mu\nu}$, 
\begin{equation}
\frac{2}{\sqrt{-g}} \nabla_\alpha \left[\sqrt{-g} f'(\mathbb{Q}) P^\alpha_{\ \mu\nu} \right] + f'(\mathbb{Q}) q_{\mu\nu} - \frac{1}{2} f(\mathbb{Q}) g_{\mu\nu} = \kappa \mathcal{T}_{\mu\nu},
\end{equation}
where $\mathcal{T}_{\mu\nu}$ is the energy-momentum tensor of matter and the prime means the derivative of $f(\mathbb{Q})$ with respect to $\mathbb{Q}$.
The non-metricity conjugate $P^\alpha{}_{\mu\nu}$ and the symmetric tensor $q_{\mu\nu}$ are defined as
\begin{equation}
 P^\alpha{}_{\mu\nu}=-\frac{1}{4}Q^\alpha{}_{\mu\nu}+\frac{1}{2}Q_{(\mu}{}^{\alpha}{}_{\nu)} 
    +\frac{1}{4}Q^\alpha g_{\mu\nu}-\frac{1}{4}\left(\overline{Q}^\alpha g_{\mu\nu}+\delta^\alpha{}_{(\mu} Q_{\nu)}\right)  ,
\end{equation}
\begin{equation}
q_{\mu\nu}=P_{(\mu|\alpha\beta}Q_{\nu)}{}^{\alpha\beta} - 2P^{\alpha\beta}{}_{(\nu} Q_{\alpha\beta|\mu)}.
\end{equation}
Similarly, varying the action with respect to the connection yields the connection field equations,
\begin{equation}
\nabla_\mu \nabla_\nu \left(\sqrt{-g} f'(\mathbb{Q}) P^{\mu\nu}_{\ \ \alpha} \right) = 0.
\end{equation}

%%%%%%%%%%%%%%%%%%%%%%%%%%%%%%%%%%%%%%%%%%%%%%%%%%%%%
\section{Static and Spherically Symmetric Solution and Pretreatment}
\label{sec:sssol}
%%%%%%%%%%%%%%%%%%%%%%%%%%%%%%%%%%%%%%%%%%%%%%%%%%%%%

We analyze a BH solution in  $f(\mathbb{Q})$  gravity and discuss the behavior of test fields in this BH background, which prepares the stage for calculating the QNMs  during the ringdown phase in the next section.

Fabio {\em et al}. \cite{DAmbrosio:2021zpm} presented a static and spherically symmetric BH solution in the following model of $f(\mathbb{Q})$  gravity,
\begin{equation}
    f(\mathbb{Q})=\mathbb{Q}+\alpha\mathbb{Q}^2+\mathcal{O}\left(\alpha^3\right),
\end{equation}
where $\alpha$ is a small parameter and it can serve as a perturbative expansion coefficient for the general form of  $f(\mathbb{Q})$. 
The corresponding metric is given by
\begin{equation}
\label{eq:metric}
    \dif s^2=-F(r)\dif t^2+\frac{1}{F(r)}\dif r^2+r^2\dif\theta^2+r^2\sin^2\theta\dif\varphi^2,
\end{equation}
with the shape function $F(r)$ expressed as
\begin{equation}
    \label{eq:original_solution}
    F(r)=1-\frac{2M+\alpha c_1+\alpha^2\left(c_2-16M^2(3c_3+c_4)\right)}{r}
    -\alpha^2\frac{48M^2 c_7}{r}\ln(r),  
\end{equation}
where $c_i$'s    $(i=1,\ldots 7)$ are integration constants in Ref.~\cite{DAmbrosio:2021zpm}, and the constants that do not appear in the metric function, i.e., $c_5$ and $c_6$, are determined to be zero when one requires that the solution goes back to the Schwarzschild case in the limit of $\alpha\to 0$. 

In the following investigations, we set $M = 1/2$, which fixes a scaling symmetry in the wave-like equation. This choice does not affect generality as the physical mass can be recovered when the relation, $\omega M = \const.$, is used  \cite{Konoplya:2011qq}. 
Owing to the arbitrariness of some constants without observable impact, we let $c_1$ be $a$, introduce scale $r_0$ by moving $c_2-16M^2(3c_3+c_4)$ as a whole into the logarithmic term, and summarize all factors outside the logarithmic term as $\alpha^2 d$. The metric function can be simplified as
\begin{equation}
    F(r)=1-\frac{1+a\alpha}{r}+\alpha^2d\,\frac{\ln(r/r_0)}{r},
\end{equation}
where $a$, $d$ and $r_0$ are independent constants that come from $c_1, c_2, c_3, c_4$ and $c_7$. To further simplify, we redefine the parameters as follows:
\begin{equation}
    \label{eq:replacement1}
    \alpha\longrightarrow\frac{\alpha}{a}, \qquad d\longrightarrow a^2k.
\end{equation}

\begin{remark}
This transformation is a rescaling of $\alpha$ and does not affect the smallness condition $a\alpha \ll 1$. The original $\alpha$ can always be restored when it is multiplied by a constant.
\end{remark}

Under the redefinition of constants, the metric function reduces to its simplest form,
\begin{equation}
\label{eq:f(r)}
F(r)=1-\frac{1}{r}-\alpha\frac{1}{r}+\alpha^2k\,\frac{\ln(r/r_0)}{r}.
\end{equation}

Now the three parameters left are mathematically independent and cannot be further simplified.

It is important to note that the above vacuum solution in  $f(\mathbb{Q})$  gravity differs from that in  $f(\mathbb{T})$  gravity \cite{Zhao:2022gxl}. 
Additionally, the logarithmic term in the metric differs from that of Einstein-Yang-Mills gravity \cite{Mazharimousavi:2008xh,Guo:2020caw}, despite a similar mathematical structure. In Sec.\ \ref{sec:comparison}, we will explore how this unique feature significantly influences the $f(\mathbb{Q})$ BH’s behavior at a moderate distance.

\begin{figure}[!ht]
	\centering
	\includegraphics[width=0.6\linewidth]{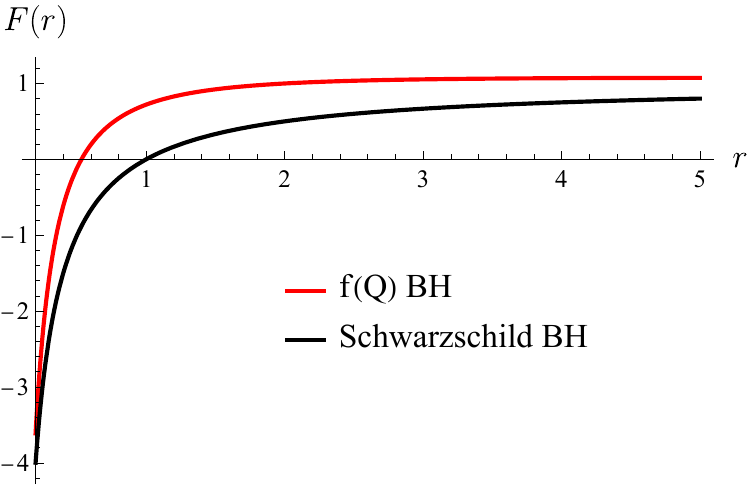}
	\captionsetup{width=.9\textwidth}
	\caption{The shape function $F(r)$ of $f(\mathbb{Q})$ BHs, where $\alpha = 0.2$, $k = 10$, and $r_0 = 0.1$ are chosen. The shape function of Schwarzschild BHs is given for comparison.}
	\label{fig:F(r)-r}
\end{figure}

Figure \ref{fig:F(r)-r} illustrates the shape function $F(r)$ of the  $f(\mathbb{Q})$  BH. 
As shown, the solution shares the same asymptotic behavior as the Schwarzschild BH: $F(r) \to 1$ as $r \to \infty$, and it diverges as $r \to 0$. For the chosen positive parameters ($r_0>0$, $k>0$ and $\alpha>0$), the $f(\mathbb{Q})$ BH features a single event horizon,
\begin{equation}
   r_{\rm H}=\alpha ^2 k\, W\!\left(\frac{r_0 \me^{\frac{\alpha +1}{\alpha ^2 k}}}{\alpha ^2 k}\right),
\end{equation} 
where $W(\cdot)$ is the Lambert W function.

To analyze waveforms, the wave function can be expressed as $\Phi(r,t)=\phi(r)\me^{-\mi\omega t}$, where $\omega$ is a complex frequency. The real part of $\omega$ represents the oscillation frequency, while the imaginary part describes the damping rate. It has been shown that the perturbation equations of massless scalar fields governed by the Klein-Gordon equation and of vector (electromagnetic) fields governed by the Maxwell equations  under the Levi-Civita connection,\footnote{The Levi-Civita connection is consistent with the coincident gauge condition,  $\Gamma^{\alpha}_{\;\;\mu\nu}=0$, 
which is employed in solving black hole solutions within both $f(\mathbb{Q})$ and $f(\mathbb{T})$ gravity frameworks. Under this gauge, the connection is expressed as
 $0=\Gamma^{\alpha}_{\;\mu\nu} = \{{}^\alpha_{\;\mu\nu}\}
+K^{\alpha}_{\;\mu\nu} + L^{\alpha}_{\;\mu\nu}$,
where $\{{}^\alpha_{\;\mu\nu}\}$ is the Levi-Civita connection, 
$K^{\alpha}_{\;\mu\nu}$ is
the contortion tensor, and $L^{\alpha}_{\;\mu\nu}$ is the disformation tensor \cite{Heisenberg:2023lru}.
Since the contortion tensor vanishes, 
$K^{\alpha}_{\;\mu\nu}=\frac{1}{2} T^{\alpha}_{\;\mu\nu} + T^{\;\;\alpha}_{(\mu\;\;\nu)}=0$
and the non-metricity becomes $Q_{\alpha\mu\nu}=\nabla_\alpha g_{\mu\nu}=\partial_\alpha g_{\mu\nu}$,
the Levi-Civita connection reduces to the disformation tensor with a negative sign $\{{}^\alpha_{\;\mu\nu}\} = - L^{\alpha}_{\;\mu\nu}$.
Substituting the expression for the disformation tensor, we have
\[
\begin{split}
\{{}^\alpha_{\;\mu\nu}\}  &= -\frac{1}{2} Q^{\alpha}_{\;\mu\nu} + Q^{\;\;\alpha}_{(\mu\;\;\nu)}\\
&= \frac{1}{2} g^{\alpha\beta}\left(\partial_{\mu} g_{\beta\nu}+\partial_{\nu}g_{\beta\mu}- \partial_{\beta} g_{\mu\nu}\right).
\end{split}
\]
This formulation highlights the direct relationship between the Levi-Civita connection and the metric’s disformation in the coincident gauge.
} can be separated~\cite{Konoplya:2019hlu,Zhao:2022gxl} into the following form,
\begin{equation}   \left(\partial^2_{r^*}+\omega^2\right)\phi(r^*)=V_{\mathrm{eff}}\ \phi(r^*),
    \label{eq:wave_equation}
\end{equation}
where $r^*$ is the tortoise coordinate defined by
\begin{equation}
\label{eq:tortoise}
    \frac{\dif r^*}{\dif r}=\frac{1}{F(r)},
\end{equation}
and $V_\mathrm{eff}$ is the effective potential given by
\begin{equation}
    \label{eq:Veff}
    V_\mathrm{eff}=F(r) \left[\frac{l(l+1)}{r^2}+\frac{1-s^2}{r}\frac{\dif F(r)}{\dif r}\right].
\end{equation}
Here, $s = 0$ corresponds to the scalar field, $s = 1$ describes the electromagnetic field, and $l$ is the orbital angular momentum of the test field. 

\begin{figure}[!ht]
     \centering
     \begin{subfigure}[b]{0.45\textwidth}
         \centering
         \includegraphics[width=\textwidth]{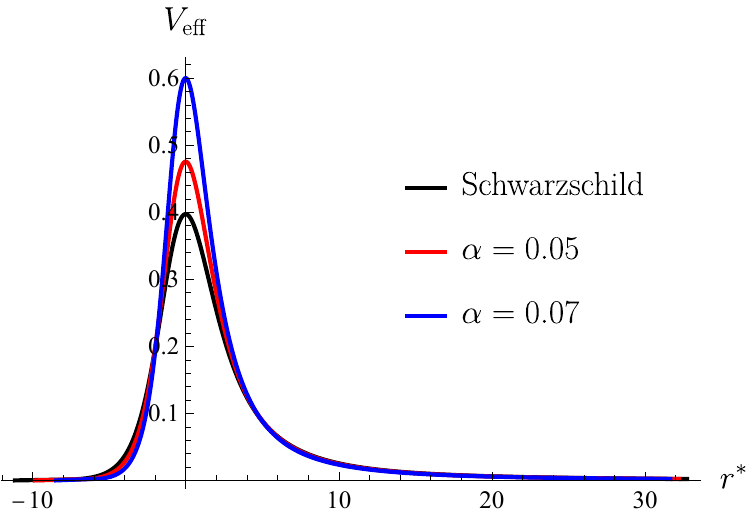}
         \caption{Scalar field perturbation.}
         %\label{fig:}
     \end{subfigure}
     %\hfill
     \begin{subfigure}[b]{0.45\textwidth}
         \centering
         \includegraphics[width=\textwidth]{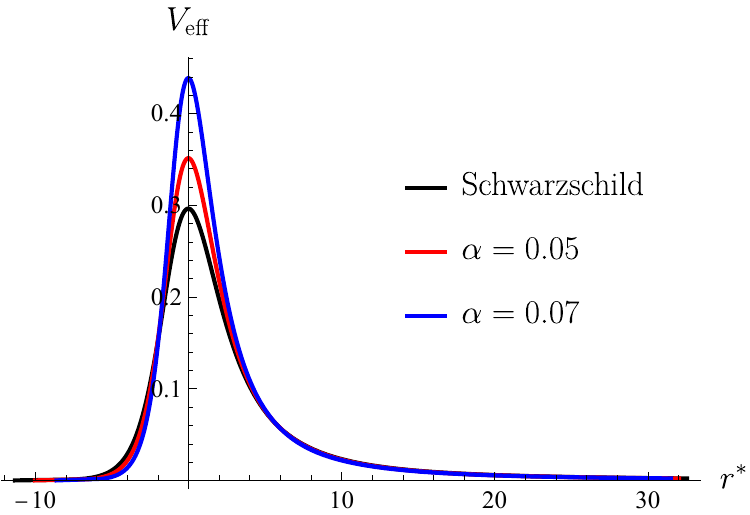}
         \caption{Electromagnetic field perturbation.}
         %\label{fig:}
     \end{subfigure}
      \captionsetup{width=.9\textwidth}
       \caption{The effective potential $V_\mathrm{eff}$  as a function of the tortoise coordinate  $r^*$, where the orbital angular momentum  $l$ is fixed to be $1$. Two sets of parameters are chosen for each type of perturbations: Set 1 ($\alpha = 0.05$,  $k = 20$,  $r_0 = 0.1$) and set 2 ($\alpha = 0.07$,  $k = 20$,  $r_0 = 0.1$). For comparison,  $V_\mathrm{eff}$  of a Schwarzschild BH is also included.}
        \label{fig:Veff}
\end{figure}

Figure \ref{fig:Veff} illustrates the behavior of  $V_\mathrm{eff}$  as a function of the tortoise coordinate  $r^*$, 
where the parameter $\alpha$ is suitably fixed for the comparison to the effective potential of a Schwarzschild BH for reference. 
The plot shows that  $V_\mathrm{eff}$  consistently scales upward, 
and its peak shifts closer to the horizon as the parameter $\alpha$ increases.
Notably, this trend is consistent in different types of perturbations. 
While we provide numerical results for each type of perturbations, 
our discussion primarily focuses on the scalar field perturbation in order to explore the properties of BHs in $f(\mathbb{Q})$  gravity. 

Moreover,  $V_\mathrm{eff}$  vanishes at both the spatial infinity ($r^* \to \infty$) and the event horizon ($r^* \to -\infty$). This behavior contrasts with that observed in  $f(\mathbb{T})$  gravity, where  $V_\mathrm{eff}$ of $f(\mathbb{T})$ approaches a nonzero positive value near the horizon (see Figure 2 in Ref.\ \cite{Zhao:2022gxl}). Such a difference typically arises in cases where the components of BH metrics, $g_{00}$  and  $g_{11}$, do not satisfy the condition:  $g_{00} g_{11} = -1$, and thus the zeros of the function  $1/g_{11}$  (definition of horizons) are not the zeros of $g_{00}$. 
The emergence of a non-vanishing effective potential is evidently associated with certain non-trivial features in the late-time behavior of gravitational waves. Notably, this phenomenon arises exclusively in a specific class of solutions within $f(\mathbb{T})$ gravity, thereby offering a potential means of distinguishing between different gravitational theories. A comprehensive analysis of this feature will be presented in Appendix, where some comments are emphasized on avoiding fake nonzero behaviors of effective potentials near horizons in any gravitational theories.

However, in our case, as described in Eq.\ \eqref{eq:metric}, the condition  $g_{00} g_{11} = -1$  holds, and the zeros of $1/g_{11}$ coincide with those of $g_{00}$. Consequently, no choice of parameters results in a nonzero effective potential near the horizon, as indicated in Eq.\ \eqref{eq:Veff}. This provides us a practical way to distinguish the solution of $f(\mathbb{Q})$ from that of $f(\mathbb{T})$ to some degree, as we will discuss in detail in Sec.\ \ref{sec:comparison}.

%%%%%%%%%%%%%%%%%%%%%%%%%%%%%%%%%%%%%%%%%%%%%%%%
\section{Method and Result}
\label{sec:numeric}
%%%%%%%%%%%%%%%%%%%%%%%%%%%%%%%%%%%%%%%%%%%%%%%%

Now we employ the finite difference method to 
compute the waveform of perturbations and  
apply the WKB approximation 
to calculate QNFs in a broad range of parameter values.

%%%%%%%%%%%%%%%%%%%%%%%%%%%%%%%%%%%%%%%%%%%%%%%%
\subsection{Finite difference method}
%%%%%%%%%%%%%%%%%%%%%%%%%%%%%%%%%%%%%%%%%%%%%%%%

\begin{figure}[!ht]
\centering
\includegraphics[width=0.6\textwidth]{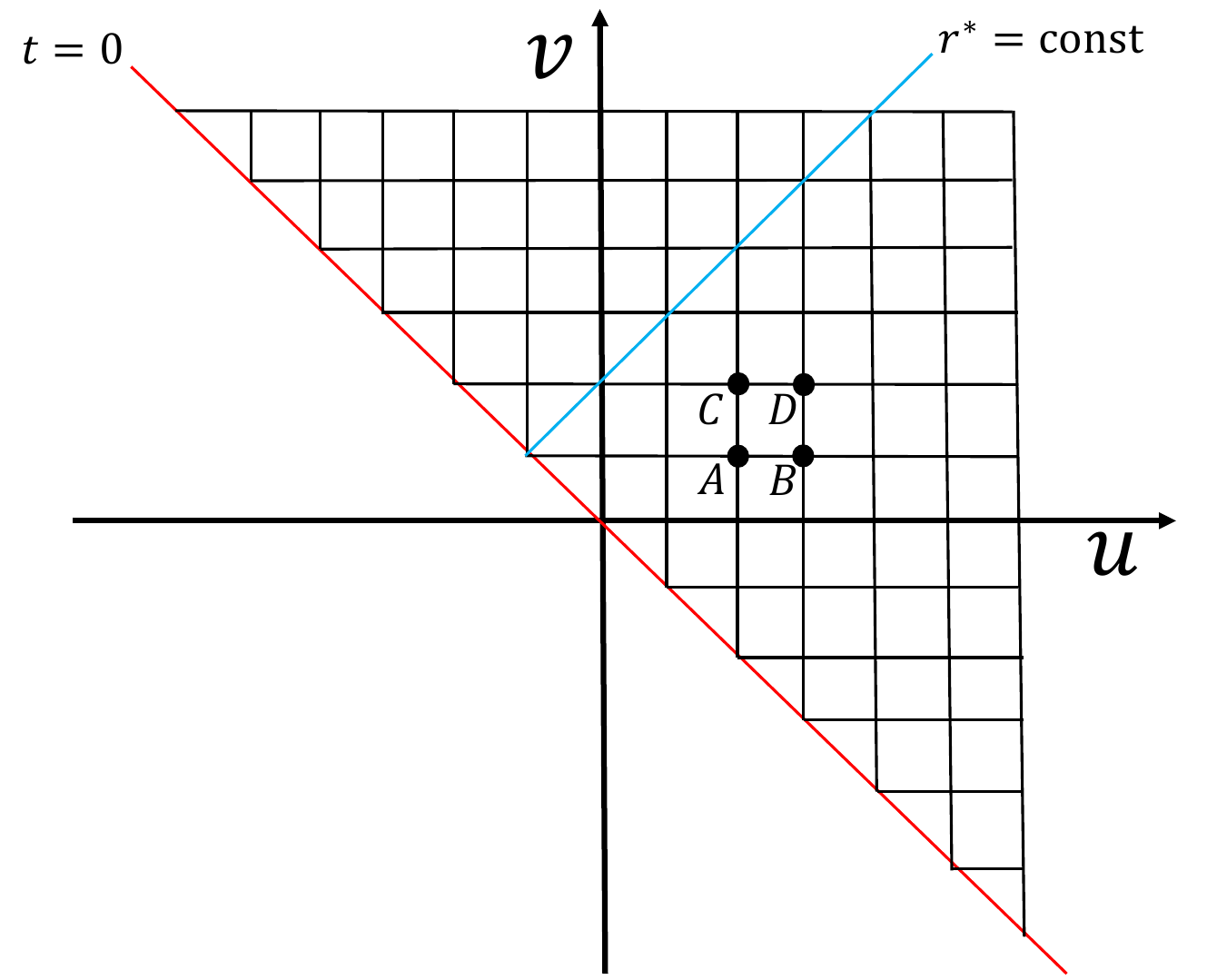}
\captionsetup{width=0.9\textwidth}
\caption{The integration grid is chosen for solving the wave equation. Each grid point corresponds to a value of the wave intensity  $\Phi(u,v)$. The points  $A$, $B$, $C$ and $D$  illustrate the relative positioning of adjacent points. Initial conditions are assigned along the  $t=0$  line, and the time decay behavior for a specific  $r^*$  is extracted along the  $r^*=\mathrm{const.}$  line.}
\label{fig:u-v_grid}
\end{figure}

The finite difference method, introduced by Gundlach et al.\ in 1994 \cite{Gundlach:1993tp},
remains widely used due to its simplicity and rigorously established convergence. 
Its stability has been fully demonstrated \cite{Konoplya:2011qq}.
This method operates in the $u-v$ coordinates, defined as
\begin{equation}
    \dif u=\dif t-\dif r^*, \qquad \dif v=\dif t+\dif r^*.
\end{equation}
To implement the method, a $u-v$ grid with a uniform spacing $h$ is constructed, see Figure~\ref{fig:u-v_grid}.

The wave equation, given in Eq.~\eqref{eq:wave_equation}, is reformulated as
\begin{equation}
    \left[4\partial_u \partial_v+V_\mathrm{eff}(u,v)\right]\Phi(u,v)=0,
\end{equation}
and is discretized into the form,
\begin{equation}
    \Phi(D)=\Phi(B)+\Phi(C)-\Phi(A)
    -\frac{h^2}{8}V_\mathrm{eff}(A)\left[\Phi(B)+\Phi(C)\right]+\mathcal{O}\left(h^4\right).
    \label{eq:integration_eq}
\end{equation}
This equation gives the basis for numerical integration.
According to Ref.~\cite{Gundlach:1993tp}, the local truncation error of this method is $\mathcal{O}(h^4)$. Considering the total number of grid points is $\mathcal{O}(h^{-2})$, the global error resulting from error propagation is $\mathcal{O}(h^2)$.

If the initial values are specified along the lower left edge of the integration grid, which stands for $t=0$, Eq.\ \eqref{eq:integration_eq} can be used iteratively to compute the values of $\Phi$ across the grid. 
By transforming back to the $t-r^*$ coordinate system, we can obtain the relation between $\Phi(t, r^*)$ and $t$ for any chosen $r^*$.

It is important to note that we need an approximate expression of the tortoise coordinate in the numerical computation of the effective potential. Specifically, an exact expression of  $r^*$ cannot be directly derived from
\begin{equation}
    r^*=\int  \frac{\dif r}{F(r)}.
\end{equation}
To address this challenge, we expand the integral near $\alpha = 0$ and numerically determine the horizon $r_\mathrm{H}$, where $F(r_\mathrm{H}) = 0$. 
By imposing the boundary condition $r^* \to -\infty$ as $r \to r_\mathrm{H}$, 
and retaining the terms up to the first-order correction in $\alpha$, 
we derive the following expression for the tortoise coordinate,
\begin{equation}
    r^*\approx r+\ln(r-2M)+\alpha\left[\frac{1}{2M-r}+\ln(r-2M)\right].\label{approxtorcor}
\end{equation}
This approximate formulation will be used in our following calculations.

\begin{figure}[!ht]
\centering
\includegraphics[width=.6\textwidth]{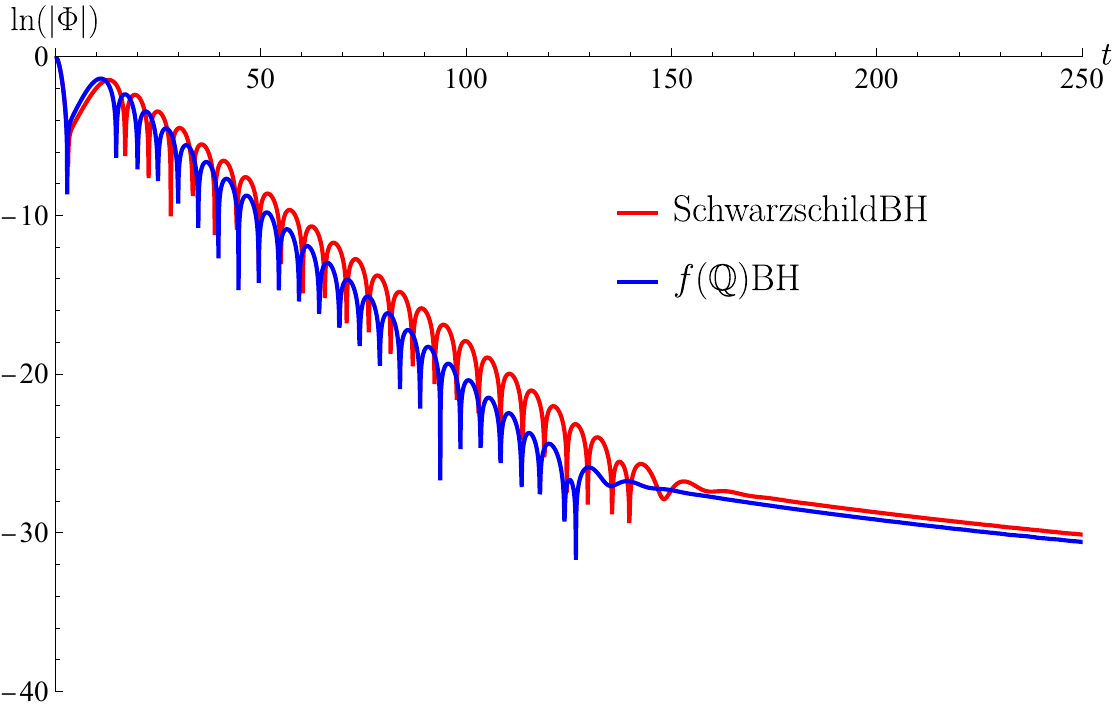}
\captionsetup{width=.9\textwidth}
\caption{The waveform of scalar field perturbations is shown with parameters $\alpha = 0.05$, $k = 20$, and $r_0 = 0.1$ in $f(\mathbb{Q})$ gravity. Here $s=0$ and $l=1$. The case of the Schwarzschild black hole is presented for comparison, also with $s=0$ and $l=1$.}
\label{fig:time_0}
\end{figure}

Figure \ref{fig:time_0} illustrates the results of exemplary numerical integrations, where $\alpha=0.05$, $k=20$, and $r_0=0.1$ are fixed, 
and $r^*$ is held constant, $r^*=5$.
For the majority of our discussions in this subsection, 
we adopt this set of illustrative parameters. %: $\alpha = 0.05$, $k = 20$, and $r_0 = 0.1$. 
It is important to clarify that these values are chosen not from observational constraints, 
which are not yet available for this specific $f(\mathbb{Q})$ model, but for two clear purposes. 
Firstly, they satisfy the theoretical validity condition of the perturbative solution, that is, 
the second-order term in $\alpha$, scaled by the factor $k$, is smaller than the first-order term, 
and both of them remain smaller than the zero-order term  in Eq.~\eqref{eq:f(r)}. 
Secondly, these specific values serve to qualitatively demonstrate the distinct influence of 
each parameter on the quasinormal modes, which is a primary goal of this study. 
The much broader parameter space will be systematically explored in Sec.\ \ref{sec:wkb}.
The scalar field perturbation is performed as a Gaussian wave packet, $\Phi(r^*)=\exp[-(r^*-r^*_c)^2/2]$, whose center is at $r^*_c=2$. 
The initial perturbation is applied as a boundary condition for numerical integration along the red line ($t=0$ fixed) in Figure ~\ref{fig:u-v_grid}.
The result of the Schwarzschild BH is included for comparison.

We can see that the waveform of scalar field perturbations around the $f(\mathbb{Q})$ BH is similar to that around the Schwarzschild BH, that is, the three-stage decay process: 
an initial phase influenced by the perturbation’s type and location, the ringdown period characterized by QNMs, 
and finally, a late-time tail phase described by a smooth power-law decay (a curve in the logarithmic plot). 
It is clear that the $f(\mathbb{Q})$ BH demonstrates the comparable early, mid-term, and late behaviors to those in the Schwarzschild BH.

The above phenomenon contrasts with that observed in $f(\mathbb{T})$ gravity for certain solutions (see Figure \ 6 in Ref.\ \cite{Zhao:2022gxl}), 
where the waveform of the extremely late stage perturbation field still remains oscillating, owing partly to the effective potential $V_\mathrm{eff}$ approaching a nonzero value near the horizon in $f(\mathbb{T})$ gravity. 
This distinction offers a potential observational means to differentiate between $f(\mathbb{Q})$ and $f(\mathbb{T})$ gravity theories.

\begin{figure}[!ht]
	\centering
	\includegraphics[width=.6\textwidth]{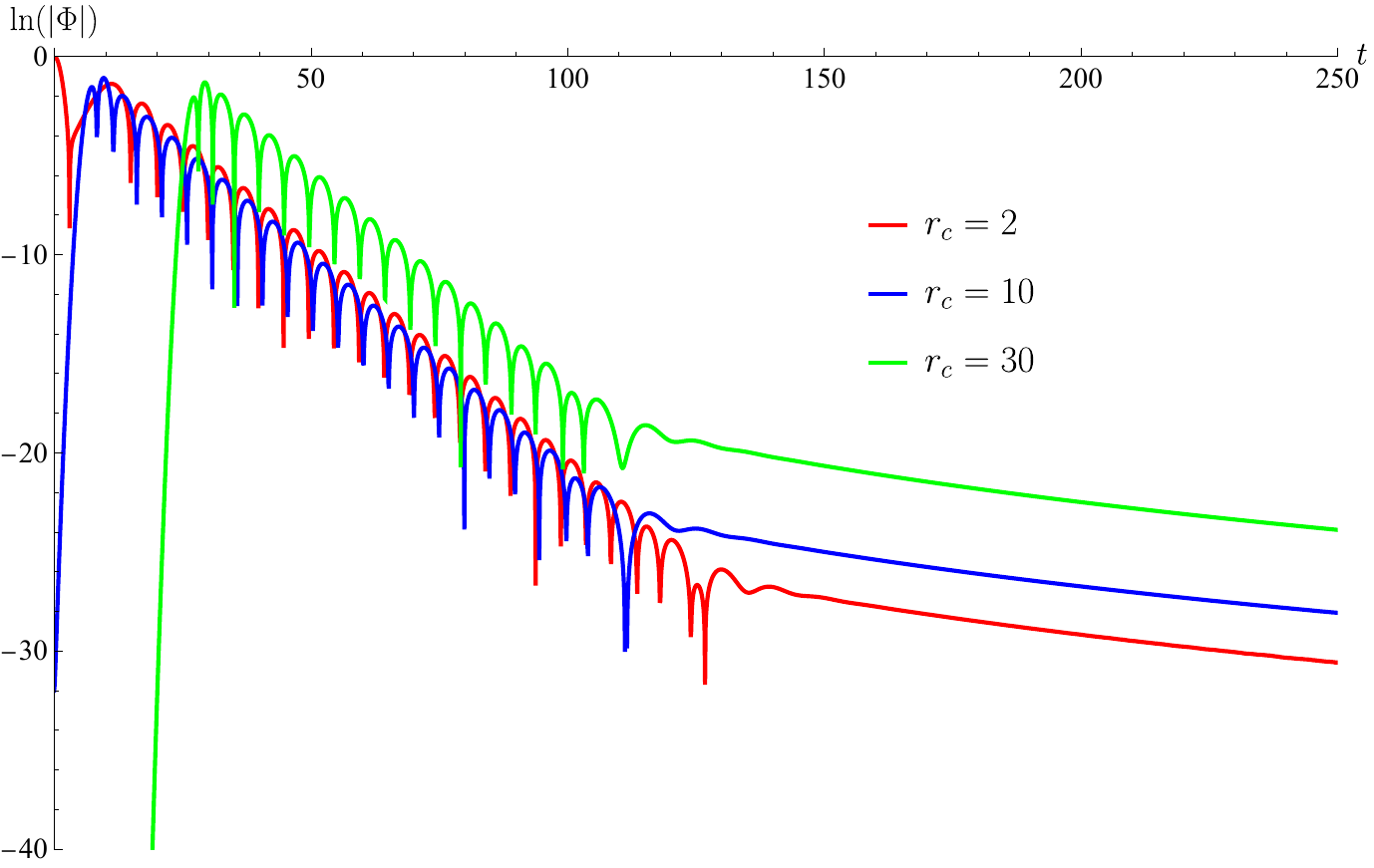}
	\captionsetup{width=.9\textwidth}
	\caption{
		Waveform of scalar field perturbations is shown when the center of perturbation wave packets, $r_c$, is varied, where $l=1$, $\alpha=0.05$, $k=20$,  and $r_0=0.1$ are set. The oscillation frequency is independent of the values of $r_c$, but the onset time of tail phases is dependent on them.}
	\label{fig:diff_rc}
\end{figure}

If the center of Gaussian wave packets is altered, the waveform takes a translational displacement along the direction of time and the onset time of tail phases changes, 
but the oscillation frequency $\omega$ keeps unchanged, as shown in Figure ~\ref{fig:diff_rc}. This behavior supports the interpretation that the oscillation frequency depends on  $V_\mathrm{eff}$, but does not on $r_c$.
QNFs can be extracted in terms of data fitting with a suitable waveform, where the fitting process can be efficiently performed with the help of the Prony method \cite{Konoplya:2011qq}. In the ringdown phase, the waveform is usually described by
\begin{equation}
    \Phi(t)=\mathrm{Re}\sum_n A_n\me^{-\mi [(a_n+\mi b_n)t-B_n]},
\end{equation}
where $n$ is the index number of a specific mode, $A_n$ is the relative intensity of the mode, $B_n$ is the initial  phase, and $a_n$ and $b_n$ stand for the real and imaginary parts of QNF of mode $n$. The summation index $n$ runs over all damped complex exponentials included in the model~\cite{Berti:2007dg}. Importantly, the range of $n$ is not fixed but rather model-dependent and  estimated from the data. The choice of the maximum value of $n$ depends on the complexity of the waveform, the signal-to-noise ratio, and the length of the fitting window. In practice, a small number of modes (typically $n$ = 1 to 3) is usually sufficient for accurate modeling. As mentioned later, we will use the Prony method to verify that the accuracy of the WKB method remains acceptable for smaller values of $l$ in the cases we consider. 

\begin{figure}[!ht]
	\centering
	\begin{subfigure}[b]{0.45\textwidth}
		\centering
		\includegraphics[width=\textwidth]{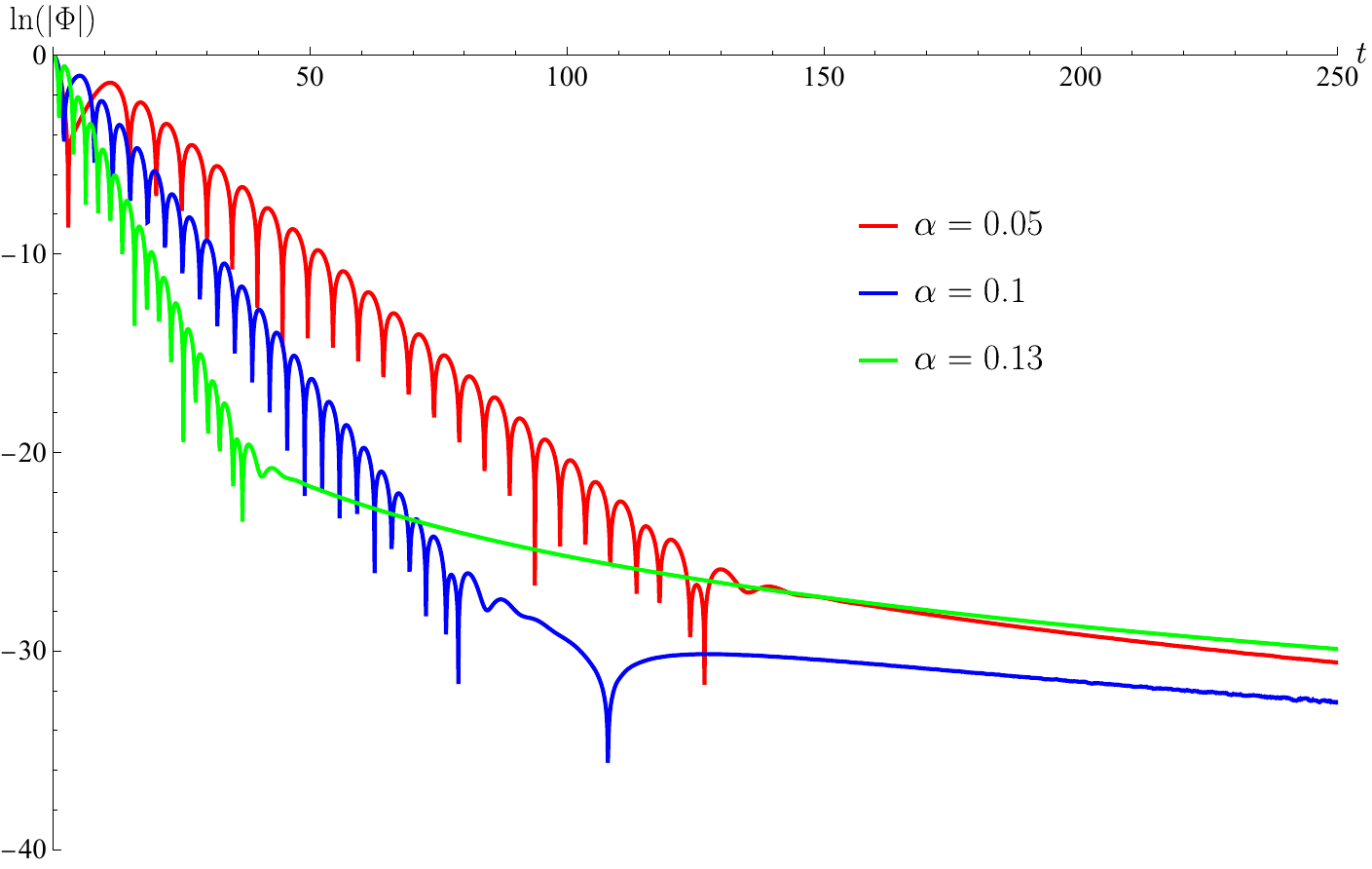}
		\caption{}
		%\label{fig:sca-per}
	\end{subfigure}
	%\hfill
	\begin{subfigure}[b]{0.45\textwidth}
		\centering
		\includegraphics[width=\textwidth]{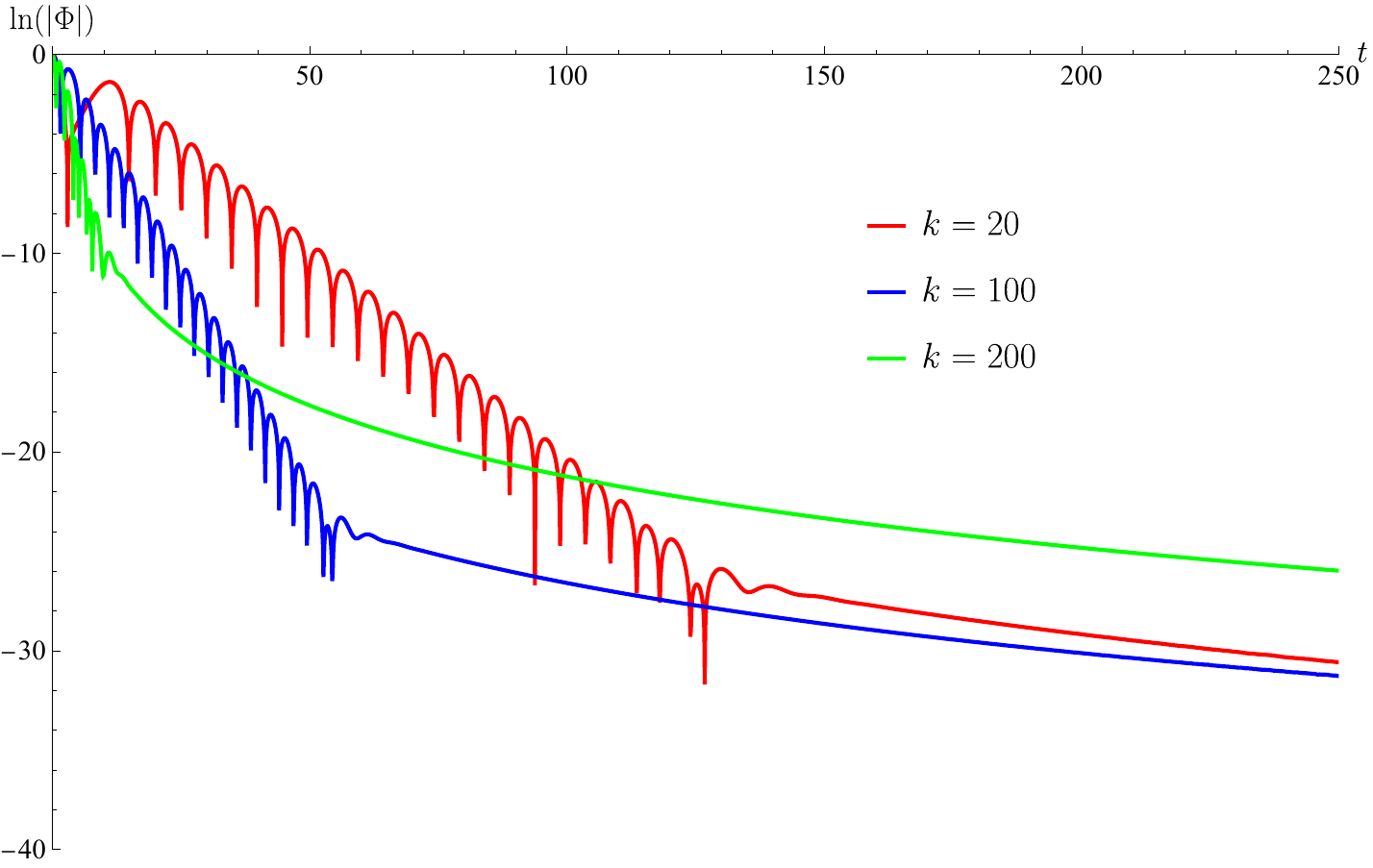}
		\caption{}
		%\label{fig:ele-per}
	\end{subfigure}
	\begin{subfigure}[b]{0.45\textwidth}
		\centering
		\includegraphics[width=\textwidth]{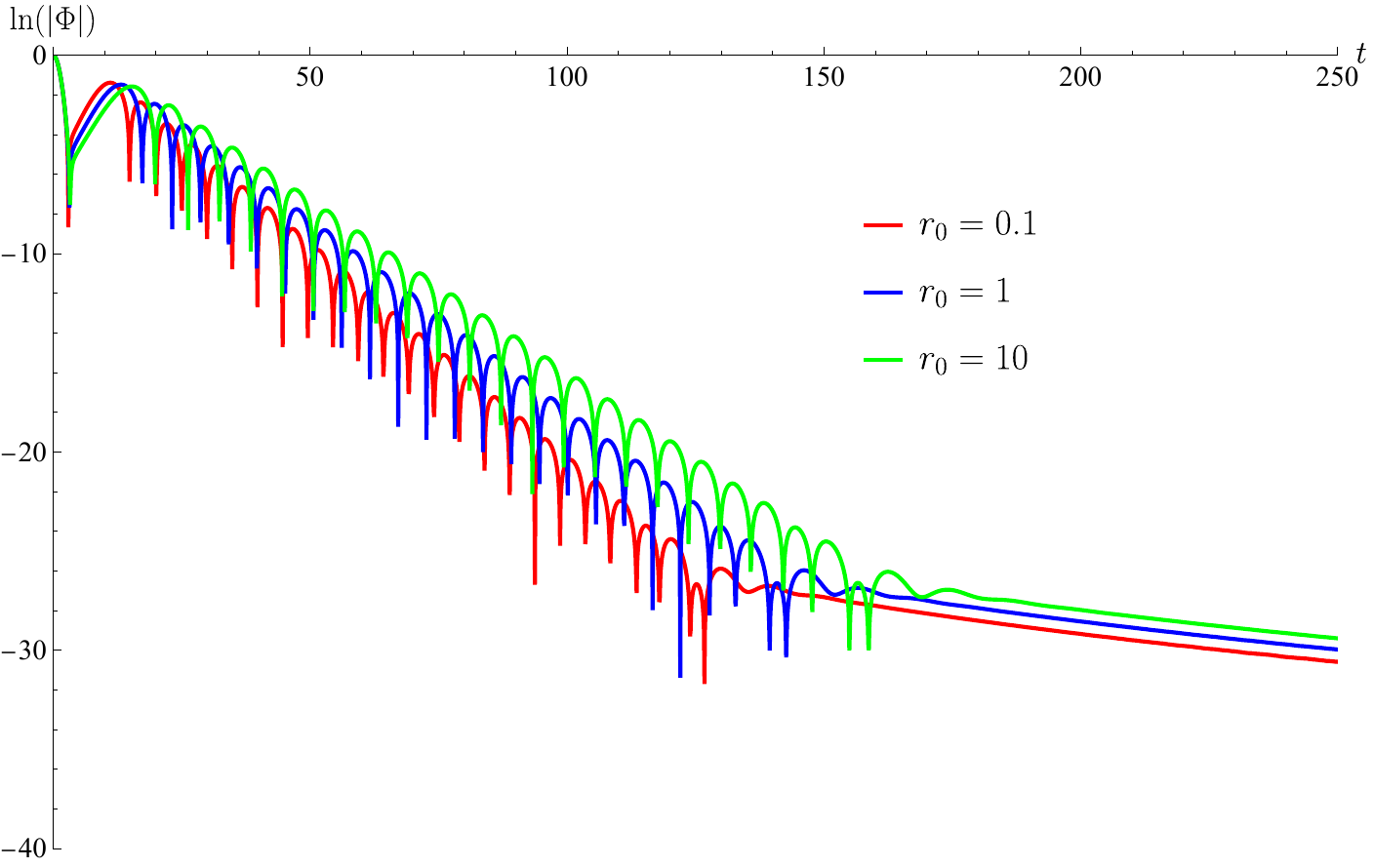}
		\caption{}
		%\label{fig:ele-per}
	\end{subfigure}
	\captionsetup{width=.9\textwidth}
	\caption{
		Waveform of scalar field perturbations under variations of the three parameters, where $l=1$ is fixed for the three cases. 
        Diagram (a) corresponds to a varying $\alpha$ but $k=20$ and $r_0=0.1$, 
        diagram (b) corresponds to a varying $k$ but $\alpha=0.05$ and $r_0=0.1$,
        and diagram (c) corresponds to a varying $r_0$ but $\alpha=0.05$ and $k=20$.}
	\label{fig:diff_parameters}
\end{figure}

Figure \ref{fig:diff_parameters} illustrates the effects of three key parameters, $\alpha$, $k$, and $r_0$, on the waveform of scalar field perturbations. As $\alpha$ increases, representing a greater deviation from GR in terms of the action, the decay rate increases and the oscillation frequency increases, too. 
For $k$, which quantifies deviations from GR at large distances from the black holes to observers, an increase also leads to higher decay rates and oscillation frequencies.
In contrast, $r_0$ behaves differently. The increase in $r_0$ leads to a synchronous decrease in vibration frequency and attenuation rate, yet this phenomenon is far less sensitive compared to the influence of the other two parameters. Higher $r_0$ also causes the tail phase to appear earlier. This is expected because the changes in $r_0$ can be interpreted as a rescaling of the black hole’s mass. This does not introduce any effects beyond $\omega M=\mathrm{const.}$
In fact, this can be seen directly from the expression of $F(r)$. Once $r_0$ is absorbed into the radial coordinate $r$, the leading-order correction corresponds to a rescaling of the mass, followed by a subleading shift in the mass. In other words, the primary contributions are effectively modifications to the black hole mass. Therefore, if we ignore the influence of different sources contributing to the black hole's mass, the impact of $r_0$ cannot be quantified.

\begin{remark}
\label{re:r0}
In order to see that the change of $r_0$ can be regarded as a scaling of the mass, we rewrite Eq.~\eqref{eq:f(r)} in the following form,
\begin{equation}
\label{eq:remarkF(r)}
F(r)=1-\frac{1+\alpha+\alpha^2k\ln(r_0)}{r}+\alpha^2k\,\frac{\ln(r)}{r}.
\end{equation}
Since the term $1/r$ is related to the significance of mass, any value of $r_0$ can be seen as a mass scaling from the case of $r_0=1$.
\end{remark}

Moreover, the tail phase exhibits weaker intensity for larger values of $\alpha$ or $k$. Notably, when any of the three parameters, $\alpha$, $k$, or $r_0$, becomes excessively large, exceeding the parameter selection criteria outlined in Sec.\ \ref{sec:sssol}, the tail phase changes to a single, sustained oscillation with a low frequency.
Beyond these observations, no other significant effects from parameter variations have been identified. 
This is consistent with expectations, as the logarithmic correction to the Schwarzschild metric must keep sufficiently small within the chosen parameter range in order to make this correction compatible with observational constraints at large distances from the black holes to observers.
Extreme parameter values are therefore of limited physical relevance. This is consistent with the underlying assumption of a small $\alpha$ in our $f(\mathbb{Q})$ BH solution, which naturally constrains the permissible range of $\alpha$.

%%%%%%%%%%%%%%%%%%%%%%%%%%%%%%%%%%%%%%%%%%%%%%%%%%%%%
\subsection{Sixth-order WKB method}
\label{sec:wkb}
%%%%%%%%%%%%%%%%%%%%%%%%%%%%%%%%%%%%%%%%%%%%%%%%%%%%%

The WKB method is widely used in calculations of the QNMs of BHs due to its efficiency and accuracy. Originally developed independently by several researchers in 1920s for solving wave equations in quantum mechanics, the method was first applied to QNM calculations by Schutz et al. in 1980s \cite{Schutz:1985km}, see also Ref.~\cite{Gogoi:2023kjt}, owing to the similarity between the perturbation equation (Eq. \eqref{eq:wave_equation}) and the Schr\"odinger equation.

To enhance its accuracy, higher-order WKB approximations have been developed \cite{Iyer:1986np,Matyjasek:2019eeu}, where one advantage is to replace the traditional Taylor expansion with the Padé approximation, see also Ref.~\cite{Konoplya:2019hlu}. 
%It is worth noting that the WKB method is generally more accurate for a higher value of $l$ \cite{Zhao:2022gxl}.
It is worth noting that the WKB method is generally more accurate when $l$ is large  \cite{Zhao:2022gxl}, but in many cases, for specific parameter regions, the WKB method can also provide very good accuracy when $l$ is small \cite{Konoplya:2019hlu}, also see the recent work \cite{Zhao:2022gxl}.

\begin{figure}[!ht]
	\centering
	\begin{subfigure}[b]{0.45\textwidth}
		\centering
		\includegraphics[width=\textwidth]{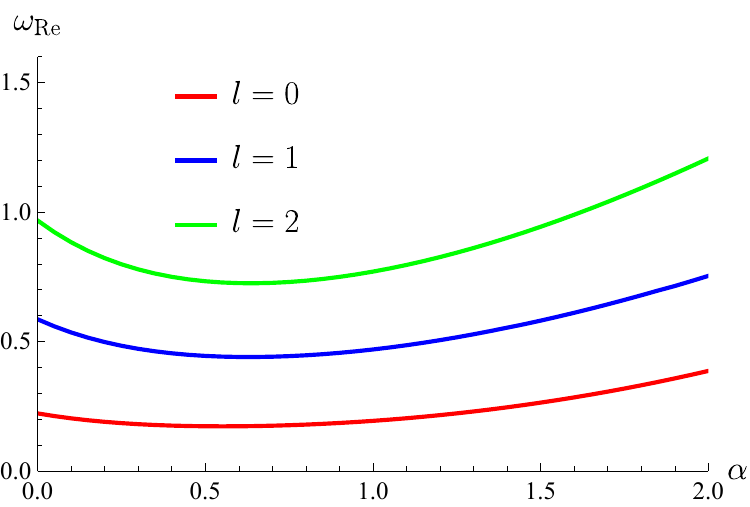}
		\caption{The real part.}
		%\label{fig:sca-per}
	\end{subfigure}
	%\hfill
	\begin{subfigure}[b]{0.45\textwidth}
		\centering
		\includegraphics[width=\textwidth]{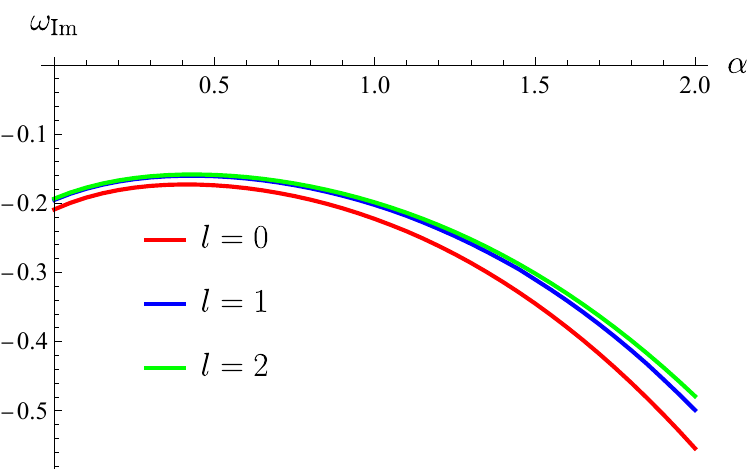}
		\caption{The imaginary part.}
		%\label{fig:ele-per}
	\end{subfigure}
	\captionsetup{width=.9\textwidth}
	\caption{The relationship of QNMs with respect to $\alpha$ in $f(\mathbb{Q})$ gravity for a varying $l$, where $k = 1$ and $r_0 = 1$ are set. The calculations are made by the sixth-order WKB method, together with the Pad\'e approximation, under a massless scalar field perturbation.}
	\label{fig:WKBdiff_alpha}
\end{figure}

\begin{figure}[!ht]
     \centering
     \begin{subfigure}[b]{0.45\textwidth}
         \centering
         \includegraphics[width=\textwidth]{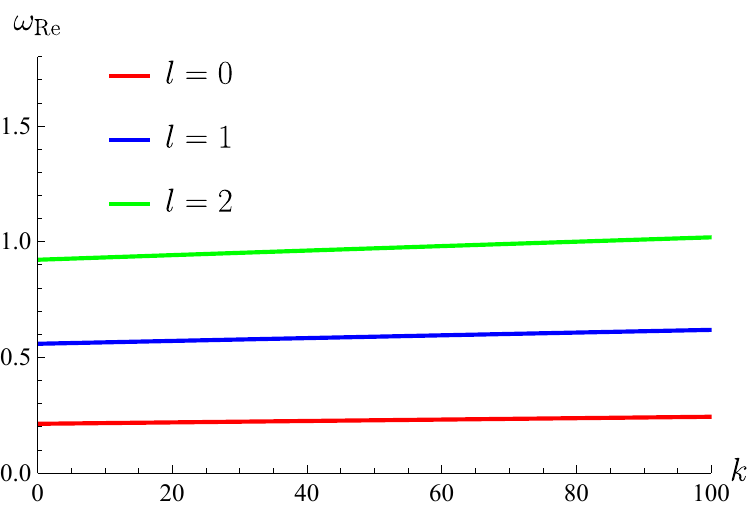}
         \caption{The real part.}
         %\label{fig:sca-per}
     \end{subfigure}
     %\hfill
     \begin{subfigure}[b]{0.45\textwidth}
         \centering
         \includegraphics[width=\textwidth]{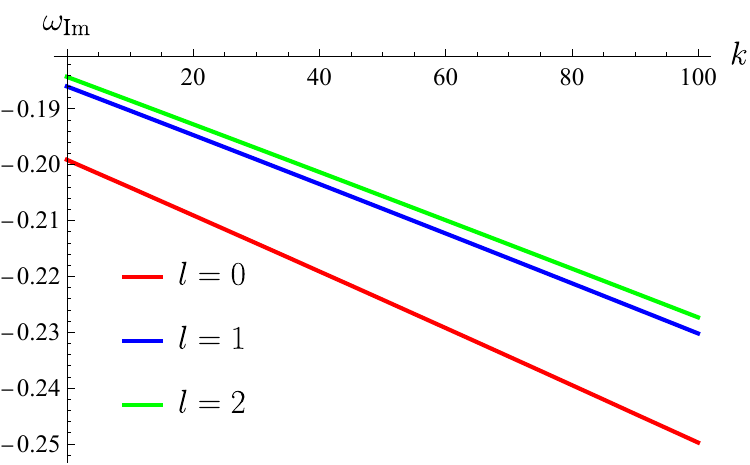}
         \caption{The imaginary part.}
         %\label{fig:ele-per}
     \end{subfigure}
      \captionsetup{width=.9\textwidth}
       \caption{The relationship of QNMs with respect to $k$ in $f(\mathbb{Q})$ gravity for a varying $l$, where $\alpha = 0.05$ and $r_0 = 1$ are set. The calculations are made by the sixth-order WKB method, together with the Pad\'e approximation, under a massless scalar field perturbation.}
        \label{fig:WKBdiff_k}
\end{figure}

\begin{figure}[!ht]
     \centering
     \begin{subfigure}[b]{0.45\textwidth}
         \centering
         \includegraphics[width=\textwidth]{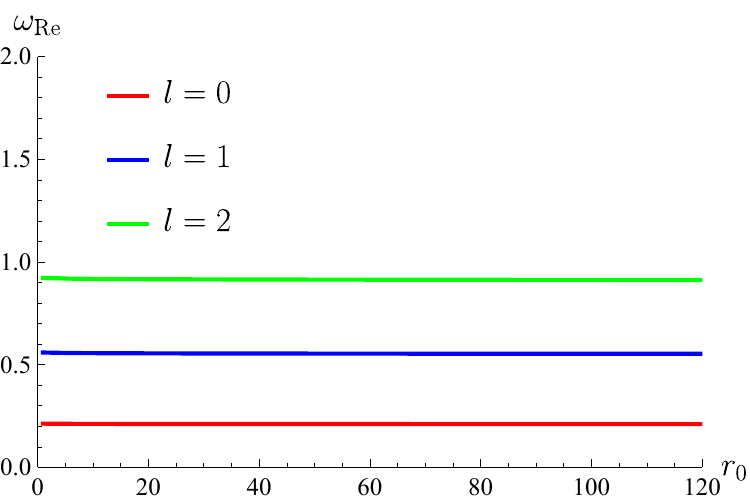}
         \caption{The real part.}
         %\label{fig:sca-per}
     \end{subfigure}
     %\hfill
     \begin{subfigure}[b]{0.45\textwidth}
         \centering
         \includegraphics[width=\textwidth]{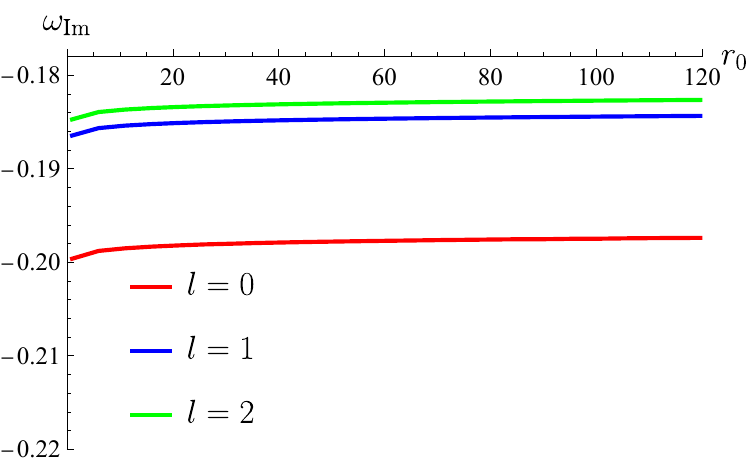}
         \caption{The imaginary part.}
         %\label{fig:ele-per}
     \end{subfigure}
      \captionsetup{width=.9\textwidth}
       \caption{The relationship of QNMs with respect to $r_0$ in $f(\mathbb{Q})$ gravity for a varying $l$, where $\alpha = 0.05$ and $k = 1$ are set. The calculations are made by the sixth-order WKB method, together with the Pad\'e approximation, under a massless scalar field perturbation.}
        \label{fig:WKBdiff_r0}
\end{figure}

Figures \ref{fig:WKBdiff_alpha}, \ref{fig:WKBdiff_k}, and \ref{fig:WKBdiff_r0} depict the relationship between $\omega$ (real and imaginary parts) and the three parameters $\alpha$, $k$, and $r_0$, respectively, for a varying angular momentum number $l$. 

For spherically symmetric BHs in  $f(\mathbb{Q})$  gravity, the real and imaginary parts of the QNM frequency $\omega$ exhibit a significant and nonlinear dependence on the parameter $\alpha$, see Figure ~\ref{fig:WKBdiff_alpha}. 
The both parts reach their minimum absolute values around $\alpha$ = 0.5 and increase as $\alpha$ diverges from this value. A bigger $\alpha$ corresponds to a higher oscillation frequency and a faster decay rate. Consequently, the relationship between the real part $\omega_\mathrm{Re}$ and imaginary part  $\omega_\mathrm{Im}$ becomes nonlinear since neither $\omega_\mathrm{Re}$ nor $\omega_\mathrm{Im}$ varies monotonically. This suggests that changes in $\alpha$ can drive the black hole to transition between different thermodynamic phases. The critical points of these phase transitions correspond to the extremal values of the real and imaginary components.  

Figure~\ref{fig:WKBdiff_k} demonstrates that the real and imaginary parts exhibit an approximately linear dependence on $k$, with their absolute values increasing as $k$ grows. That is, a bigger $k$ results in a higher oscillation frequency and a faster decay rate.

The impact of $r_0$ can be seen in Figure ~\ref{fig:WKBdiff_r0}, where any $r_0$ can effectively be redefined to be $r_0 \equiv 1$ and such an operation gives rise to an additional term proportional to $1/r$. This redefinition allows $r_0$ to be treated purely as a mass correction term, see the above Remark \ref{re:r0}.

The detailed numerical results, for the relationship of QNMs with respect to $\alpha$ in $f(\mathbb{Q})$ gravity, see Figure~\ref{fig:WKBdiff_alpha}, are presented in Table~\ref{Tab:1} for scalar field perturbations and Table~\ref{Tab:2} for vector field perturbations, respectively. 
The calculation of errors are performed by the method provided in Ref.\ \cite{konoplya2019higher}, $\Delta_k=|\omega_{k+1}-\omega_{k-1}|/2$.
Since our primary focus is on the effects of deviations from GR, characterized by $\alpha$, we present the data with $\alpha$ and $l$ as the variables. To demonstrate the accuracy of the WKB method (especially for smaller values of $l$), we choose specific parameters and give the numerical results for the waveforms using the finite difference method, and then extract the QNFs using the Prony method. Using the Prony method as the reference, we examine the relative errors of the WKB method. Specifically, we consider the first two orders of $l$ (since the WKB method generally performs better for a large $l$) for the case of $n=0$. The results are presented in Table \ref{Tab:3} and Table \ref{Tab:4}, in which the absolute values of the relative errors are mostly below 1\%, and the relative error with maximum absolute value is -5.372\%. This indicates that the WKB method is highly suitable for calculating the QNFs in our scenario.

\begin{table}[!ht]
    \centering
    \begin{adjustbox}{width=\textwidth}
    \begin{tabular}{|c|c|cc|c|cc|c|cc|c|}
    \hline
    \multirow{2}[0]{*}{$\alpha$} & \multicolumn{1}{c|}{\multirow{2}[0]{*}{$l$}} & \multicolumn{2}{c}{$n=0$} & \multicolumn{1}{|c|}{\multirow{2}[0]{*}{errors}} & \multicolumn{2}{c}{$n=1$} & \multicolumn{1}{|c|}{\multirow{2}[0]{*}{errors}} & \multicolumn{2}{c}{$n=2$} & \multicolumn{1}{|c|}{\multirow{2}[0]{*}{errors}}\\
    \cline{3-4}\cline{6-7}\cline{9-10}
          &       & \multicolumn{1}{c}{$\omega_\mathrm{Re}$} & \multicolumn{1}{c|}{$\omega_\mathrm{Im}$} & & \multicolumn{1}{c}{$\omega_\mathrm{Re}$} & \multicolumn{1}{c|}{$\omega_\mathrm{Im}$} &  & \multicolumn{1}{c}{$\omega_\mathrm{Re}$} & \multicolumn{1}{c|}{$\omega_\mathrm{Im}$}  & \\
    \hline
    \multirow{3}[0]{*}{0} & 0     & 0.222620 & -0.209111 & $1.2E{-3}$ & 0.175050 & -0.710126  & $7.2E{-3}$ & 0.156048 & -1.22490 & $4.7E{-3}$\\
          & 1     & 0.585861 & -0.195322 & $2.8E{-6}$ & 0.528913 & -0.613014  & $3.7E{-5}$& 0.457699 & -1.08433 & $6.1E{-4}$\\
          & 2     & 0.967287 & -0.193518 & $1.2E{-6}$ & 0.927692 & -0.591251  & $1.9E{-6}$ & 0.860659 & -1.01733 & $1.0E{-4}$\\
    \hline
    \multirow{3}[0]{*}{0.01} & 0     & 0.220429 & -0.207065 & $1.2E{-3}$ & 0.173323 & -0.703167  & $7.1E{-3}$ & 0.154479 & -1.21292 & $4.6E{-3}$\\
          & 1     & 0.580085 & -0.193406 & $2.8E{-6}$ & 0.523691 & -0.607004  & $3.7E{-5}$ & 0.453175 & -1.07371 & $6.0E{-4}$\\
          & 2     & 0.957750 & -0.191620 & $1.2E{-6}$ & 0.918540 & -0.585452  & $1.9E{-6}$& 0.852160 & -1.00735 & $1.0E{-4}$\\
    \hline
    \multirow{3}[0]{*}{0.02} & 0     & 0.218304 & -0.205088 & $1.2E{-3}$ & 0.171657 & -0.696493  & $7.0E{-3}$ & 0.153056 & -1.20136 & $4.6E{-3}$\\
          & 1     & 0.574473 & -0.191563 & $2.8E{-6}$ & 0.518605 & -0.601230  & $3.6E{-5}$ & 0.448753 & -1.06351 & $6.0E{-4}$\\
          & 2     & 0.948481 & -0.189793 & $1.2E{-6}$ & 0.909636 & -0.579875  & $1.9E{-6}$ & 0.843875 & -0.997769  & $1.0E{-4}$ \\
    \hline
    \multirow{3}[0]{*}{0.05} & 0     & 0.212324 & -0.199633 & $1.1E{-3}$ & 0.166947 & -0.678074 & $6.8E{-3}$ & 0.148942 & -1.16952 & $4.4E{-3}$\\
          & 1     & 0.558586 & -0.186453 & $2.8E{-6}$ & 0.504134 & -0.585252 & $3.6E{-5}$ & 0.436101 & -1.03536 & $5.9E{-4}$\\
          & 2     & 0.922233 & -0.184726 & $1.1E{-6}$ & 0.884364 & -0.564417  & $1.8E{-6}$ & 0.820268 & -0.971238  & $1.0E{-4}$\\
    \hline
    \multirow{3}[0]{*}{0.1} & 0     & 0.203579 & -0.191982 & $1.1E{-3}$ & 0.160021 & -0.652268 & $6.6E{-3}$&  0.142708 & -1.12498 & $4.3E{-3}$\\
          & 1     & 0.535095 & -0.179219  & $2.9E{-6}$ & 0.482515 & -0.562740 & $3.5E{-5}$ &  0.416983 & -0.995880  & $5.8E{-4}$\\
          & 2     & 0.883392 & -0.177542  & $1.1E{-6}$ & 0.846797 & -0.542545 & $1.8E{-6}$ &  0.784901 & -0.933827  & $9.8E{-5}$\\
    \hline
    \multirow{3}[0]{*}{0.2} & 0     & 0.190084 & -0.181137  & $1.0E{-3}$ & 0.149307 & -0.616157 & $6.3E{-3}$ & 0.133293 & -1.06238 & $4.2E{-3}$\\
          & 1     & 0.497991 & -0.168834  & $3.9E{-6}$ & 0.447648 & -0.530807 & $3.4E{-5}$ & 0.385478 & -0.940565  & $6.0E{-4}$\\
          & 2     & 0.821945 & -0.167199  & $1.2E{-6}$ & 0.786809 & -0.511208  & $1.3E{-6}$ & 0.727535 & -0.880661  & $9.9E{-5}$\\
    \hline
    \multirow{3}[0]{*}{0.5} & 0     & 0.172906 & -0.174245 & $9.6E{-4}$ &  0.136059 & -0.595802 & $6.5E{-3}$ &  0.122303 & -1.02535 & $4.4E{-3}$\\
          & 1     & 0.444467 & -0.161095  & $7.9E{-6}$& 0.392410 & -0.510205  & $4.7E{-5}$ & 0.331746 & -0.910283  & $8.9E{-4}$\\
          & 2     & 0.732588 & -0.159220  &$1.8E{-6}$ & 0.695599 & -0.488351 & $3.4E{-6}$&  0.634230 & -0.845612  & $1.5E{-4}$\\
    \hline
    \multirow{3}[0]{*}{1} & 0     & 0.193755 & -0.222795  & $7.9E{-4}$ & 0.157846 & -0.764374  & $8.1E{-3}$ & 0.145115 & -1.30943 & $6.3E{-3}$\\
          & 1     & 0.469260 & -0.202566  & $4.1E{-5}$ & 0.393509 & -0.654413  & $3.1E{-4}$ & 0.322865 & -1.18494 & $1.7E{-3}$\\
          & 2     & 0.769767 & -0.198892  & $5.4E{-6}$ & 0.712150 & -0.616491  & $3.7E{-5}$ & 0.622852 & -1.08434 & $5.9E{-4}$\\
    \hline
    \multirow{3}[0]{*}{2} & 0     & 0.386461 & -0.554405  & $2.0E{-3}$ & 0.325143 & -1.85510 & $2.0E{-3}$& 0.334315 & -3.18142 & $1.4E{-2}$\\
          & 1     & 0.753196 & -0.499051  & $1.9E{-3}$ & 0.556084 & -1.70630 & $1.5E{-2}$ & 0.480756 & -3.02778 & $3.3E{-3}$\\
          & 2     & 1.20638& -0.478970  & $2.4E{-5}$ & 1.01556& -1.54688 & $3.0E{-4}$ & 0.824896 & -2.82105 & $9.4E{-3}$\\
    \hline
    \end{tabular}
    \end{adjustbox}
    \caption{QNMs for scalar field perturbations, where $k=1$ and $r_0=1$ are fixed.}
    \label{Tab:1}
\end{table}

%\clearpage

\begin{table}[!ht]
    \centering
      \begin{adjustbox}{width=\textwidth}
    \begin{tabular}{|c|c|cc|c|cc|c|cc|c|}
    \hline
    %\multirow{2}[0]{*}{$\alpha$} & \multicolumn{1}{|c|}{\multirow{2}[0]{*}{$l$}} & \multicolumn{2}{|c|}{$n=0$} & \multicolumn{2}{|c|}{$n=1$} & \multicolumn{2}{|c|}{$n=2$} \\
    %\cline{3-8}
    %      &       & \multicolumn{1}{c}{$\omega_\mathrm{Re}$} & \multicolumn{1}{c|}{$\omega_\mathrm{Im}$} & \multicolumn{1}{c}{$\omega_\mathrm{Re}$} & \multicolumn{1}{c|}{$\omega_\mathrm{Im}$} & \multicolumn{1}{c}{$\omega_\mathrm{Re}$} & \multicolumn{1}{c|}{$\omega_\mathrm{Im}$} \\
    \multirow{2}[0]{*}{$\alpha$} & \multicolumn{1}{c|}{\multirow{2}[0]{*}{$l$}} & \multicolumn{2}{c}{$n=0$} & \multicolumn{1}{|c|}{\multirow{2}[0]{*}{errors}} & \multicolumn{2}{c}{$n=1$} & \multicolumn{1}{|c|}{\multirow{2}[0]{*}{errors}} & \multicolumn{2}{c}{$n=2$} & \multicolumn{1}{|c|}{\multirow{2}[0]{*}{errors}}\\
    \cline{3-4}\cline{6-7}\cline{9-10}
          &       & \multicolumn{1}{c}{$\omega_\mathrm{Re}$} & \multicolumn{1}{c|}{$\omega_\mathrm{Im}$} &  & \multicolumn{1}{c}{$\omega_\mathrm{Re}$} & \multicolumn{1}{c|}{$\omega_\mathrm{Im}$} &  & \multicolumn{1}{c}{$\omega_\mathrm{Re}$} & \multicolumn{1}{c|}{$\omega_\mathrm{Im}$}  & \\
    \hline
    \multirow{3}[0]{*}{0} & 1     & 0.496503  & -0.184969   & $1.8E{-5}$ & 0.428543  & -0.588209   & $5.5E{-5}$ & 0.346032  & -1.05977  & $1.8E{-3}$\\
          & 2     & 0.915190  & -0.190010   &  $1.1E{-6}$ & 0.873066  & -0.581454  & $2.5E{-6}$  & 0.801716  & -1.00339  & $1.1E{-4}$\\
          & 3     & 1.31380& -0.191233   & $2.0E{-7}$ & 1.28347& -0.579461   & $3.3E{-7}$ & 1.22756& -0.984115   &  $1.6E{-5}$\\
    \hline
    \multirow{3}[0]{*}{0.01} & 1     & 0.491602  & -0.183154  & $1.8E{-5}$ & 0.424303  & -0.582439  & $5.4E{-5}$ & 0.342598  & -1.04939 & $1.8E{-3}$\\
          & 2     & 0.906163  & -0.188145  & $1.1E{-6}$ & 0.864449  & -0.575750  & $2.5E{-6}$ & 0.793794  & -0.993551  & $1.1E{-4}$ \\
          & 3     & 1.30084& -0.189356  & $1.9E{-7}$ & 1.27081& -0.573776  & $3.3E{-7}$ & 1.21545& -0.974462  & $1.6E{-5}$ \\
    \hline
    \multirow{3}[0]{*}{0.02} & 1     & 0.486829  & -0.181404  & $1.8E{-5}$ & 0.420156  & -0.576889  & $5.4E{-5}$ & 0.339218  & -1.03941 & $1.7E{-3}$\\
          & 2     & 0.897383  & -0.186350  & $1.1E{-6}$ & 0.856057  & -0.570262  & $2.5E{-6}$ & 0.786060  & -0.984092   & $1.1E{-4}$\\
          & 3     & 1.28824& -0.187550  & $1.9E{-7}$ & 1.25849& -0.568306  & $3.3E{-7}$ & 1.20364& -0.965179  & $1.6E{-5}$ \\
    \hline
    \multirow{3}[0]{*}{0.05} & 1     & 0.473255  & -0.176539  & $1.7E{-5}$ & 0.408260  & -0.561496  & $5.2E{-5}$ & 0.329405  & -1.01185 & $1.7E{-3}$\\
          & 2     & 0.872485  & -0.181366  & $1.1E{-6}$ & 0.832195  & -0.555034  & $2.4E{-6}$ & 0.763965  & -0.957891  &  $1.1E{-4}$\\
          & 3     & 1.25254& -0.182537   & $1.9E{-7}$& 1.22354& -0.553127  & $3.2E{-7}$ & 1.17006& -0.939441   & $1.6E{-5}$\\
    \hline
    \multirow{3}[0]{*}{0.1} & 1     & 0.452992  & -0.169604  & $1.7E{-5}$ & 0.390195  & -0.539696  & $5.1E{-5}$ & 0.314156  & -0.973139  & $1.7E{-3}$ \\
          & 2     & 0.835526  & -0.174284   & $1.1E{-6}$& 0.796587  & -0.533446  & $2.4E{-6}$ & 0.730682  & -0.920881  & $1.0E{-4}$ \\
          & 3     & 1.19962& -0.175419  & $1.8E{-7}$ & 1.17158& -0.531600  & $3.1E{-7}$ & 1.11991& -0.90301 & $1.5E{-5}$\\
    \hline
    \multirow{3}[0]{*}{0.2} & 1     & 0.420362  & -0.159482  & $1.7E{-5}$ & 0.360104  & -0.508376  & $5.1E{-5}$ & 0.287673  & -0.918624  &  $1.7E{-3}$\\
          & 2     & 0.776690  & -0.164032  & $1.1E{-6}$ & 0.739284  & -0.502357  & $2.4E{-6}$ & 0.676113  & -0.868067  & $1.1E{-4}$ \\
          & 3     & 1.11559& -0.165133  & $1.8E{-7}$ & 1.08865& -0.500574   &$3.1E{-7}$ & 1.03905& -0.850773  & $1.6E{-5}$ \\
    \hline
    \multirow{3}[0]{*}{0.5} & 1     & 0.368757  & -0.150491  & $2.5E{-5}$ & 0.305531  & -0.484782  & $6.6E{-5}$ & 0.232783  & -0.887315  & $2.4E{-3}$ \\
          & 2     & 0.688460  & -0.155641   &$1.5E{-6}$ & 0.648948  & -0.478314  & $3.1E{-6}$ & 0.583125  & -0.831339  & $1.6E{-4}$ \\
          & 3     & 0.991238  & -0.156876  & $2.3E{-7}$ & 0.962735  & -0.476367  & $4.2E{-7}$ & 0.910587  & -0.812247  & $2.4E{-5}$ \\
    \hline
    \multirow{3}[0]{*}{1} & 1     & 0.367154  & -0.182131  & $5.4E{-5}$ & 0.269478  & -0.608092  & $1.8E{-4}$ & 0.173153  & -1.15792 & $8.4E{-3}$\\
          & 2     & 0.710222  & -0.192072  & $4.9E{-6}$ & 0.647785  & -0.597223  & $9.3E{-6}$ & 0.549026  & -1.05740 & $6.9E{-4}$\\
          & 3     & 1.03086& -0.194393   &$7.3E{-7}$ & 0.985534  & -0.593752  & $1.27E{-6}$ & 0.904682  & -1.02320 & $1.2E{-4}$\\
    \hline
    \multirow{3}[0]{*}{2} & 1     & 0.442449  & -0.371429  & $6.8E{-4}$ & 0.107266  & -1.47556 & $3.4E{-3}$ & 0.265593  & 0.706881   & 2.6\\
          & 2     & 1.01997& -0.436967  & $7.2E{-5}$ & 0.788724  & -1.43460 & $2.1E{-4}$ & 0.503217  & -2.71663 & $1.8E{-2}$\\
          & 3     & 1.53931& -0.451478  & $1.4E{-5}$ & 1.36606& -1.41679 & $3.1E{-5}$ & 1.09612& -2.54715 & $3.8E{-3}$\\
    \hline
    \end{tabular}
     \end{adjustbox}
    \caption{QNMs for vector field perturbations, where $k=1$ and $r_0=1$ are fixed.}
    \label{Tab:2}
\end{table}

%\clearpage

\begin{table}[htbp]
    \centering
    \begin{tabular}{|c|c|cc|cc|cc|}
    \hline
    \multicolumn{8}{|c|}{s=0} \\
    \hline
    \multirow{2}[0]{*}{$\alpha$} & \multicolumn{1}{c|}{\multirow{2}[0]{*}{$l$}} & \multicolumn{2}{c|}{WKB} & \multicolumn{2}{c|}{Prony} & \multicolumn{2}{c|}{Relative Error} \\
    \cline{3-8}
          &       & \multicolumn{1}{c|}{Re} & \multicolumn{1}{c|}{Im} & \multicolumn{1}{c|}{Re} & \multicolumn{1}{c|}{Im} & \multicolumn{1}{c|}{Re} & \multicolumn{1}{c|}{Im} \\
    \hline
    \multirow{2}[0]{*}{0} & 0     & 0.222620 & -0.209111 & 0.2247 & -0.2078 & -0.926\% & 0.631\% \\
          & 1     & 0.585861 & -0.195322 & 0.5871 & -0.1931 & -0.211\% & 1.151\% \\
    \hline
    \multirow{2}[0]{*}{0.01} & 0     & 0.220429 & -0.207065 & 0.2225 & -0.2059 & -0.931\% & 0.566\% \\
          & 1     & 0.580085 & -0.193406 & 0.5813 & -0.1913 & -0.209\% & 1.101\% \\
    \hline
    \multirow{2}[0]{*}{0.02} & 0     & 0.218304 & -0.205088 & 0.2203 & -0.2041 & -0.906\% & 0.484\% \\
          & 1     & 0.574473 & -0.191563 & 0.5756 & -0.1895 & -0.196\% & 1.089\% \\
    \hline
    \multirow{2}[0]{*}{0.05} & 0     & 0.212324 & -0.199633 & 0.2140 & -0.1992 & -0.783\% & 0.217\% \\
          & 1     & 0.558586 & -0.186453 & 0.5597 & -0.1845 & -0.199\% & 1.059\% \\
    \hline
    \multirow{2}[0]{*}{0.1} & 0     & 0.203579 & -0.191982 & 0.2044 & -0.1916 & -0.402\% & 0.199\% \\
          & 1     & 0.535095 & -0.179219 & 0.5361 & -0.1774 & -0.187\% & 1.025\% \\
    \hline
    \multirow{2}[0]{*}{0.2} & 0     & 0.190084 & -0.181137 & 0.1904 & -0.1806 & -0.166\% & 0.297\% \\
          & 1     & 0.497991 & -0.168834 & 0.4989 & -0.1672 & -0.182\% & 0.977\% \\
    \hline
    \multirow{2}[0]{*}{0.5} & 0     & 0.172906 & -0.174245 & 0.1731 & -0.1747 & -0.112\% & -0.260\% \\
          & 1     & 0.444467 & -0.161095 & 0.4454 & -0.1596 & -0.209\% & 0.937\% \\
    \hline
    \multirow{2}[0]{*}{1} & 0     & 0.193755 & -0.222795 & 0.1965 & -0.2234 & -1.397\% & -0.271\% \\
          & 1     & 0.469260 & -0.202566 & 0.4708 & -0.2003 & -0.327\% & 1.131\% \\
    \hline
    \multirow{2}[0]{*}{2} & 0     & 0.386461 & -0.554405 & 0.4084 & -0.5529 & -5.372\% & 0.272\% \\
          & 1     & 0.753196 & -0.499051 & 0.7677 & -0.4863 & -1.889\% & 2.622\% \\
    \hline
    \end{tabular}
    \caption{Error comparison between the WKB method and the Prony method for scalar field perturbations. The relative error is calculated using the Prony method as the benchmark.}
    \label{Tab:3}
\end{table}

%\clearpage

\begin{table}[htbp]
    \centering
    \begin{tabular}{|c|c|cc|cc|cc|}
    \hline
    \multicolumn{8}{|c|}{s=1} \\
    \hline
    \multirow{2}[0]{*}{$\alpha$} & \multicolumn{1}{c|}{\multirow{2}[0]{*}{$l$}} & \multicolumn{2}{c|}{WKB} & \multicolumn{2}{c|}{Prony} & \multicolumn{2}{c|}{Relative Error} \\
    \cline{3-8}
          &       & \multicolumn{1}{c|}{Re} & \multicolumn{1}{c|}{Im} & \multicolumn{1}{c|}{Re} & \multicolumn{1}{c|}{Im} & \multicolumn{1}{c|}{Re} & \multicolumn{1}{c|}{Im} \\
    \hline
    \multirow{2}[0]{*}{0} & 1     & 0.496503 & -0.184969 & 0.4977& -0.1830 & -0.241\% & 1.076\% \\
          & 2     & 0.915190 & -0.190010& 0.9158 & -0.1880 & -0.067\% & 1.069\% \\
    \hline
    \multirow{2}[0]{*}{0.01} & 1     & 0.491602 & -0.183154 & 0.4928 & -0.1813 & -0.243\% & 1.023\% \\
          & 2     & 0.906163 & -0.188145 & 0.9068 & -0.1862 & -0.070\% & 1.045\% \\
    \hline
    \multirow{2}[0]{*}{0.02} & 1     & 0.486829 & -0.181404 & 0.4880 & -0.1796 & -0.240\% & 1.004\% \\
          & 2     & 0.897383 & -0.186350& 0.8979 & -0.1844 & -0.058\% & 1.057\% \\
    \hline
    \multirow{2}[0]{*}{0.05} & 1     & 0.473255 & -0.176539 & 0.4744 & -0.1748 & -0.241\% & 0.995\% \\
          & 2     & 0.872485 & -0.181366 & 0.8729 & -0.1797 & -0.048\% & 0.927\% \\
    \hline
    \multirow{2}[0]{*}{0.1} & 1     & 0.452992 & -0.169604 & 0.4540 & -0.1680 & -0.222\% & 0.955\% \\
          & 2     & 0.835526 & -0.174284 & 0.8355 & -0.1729 & 0.003\% & 0.800\%\\
    \hline
    \multirow{2}[0]{*}{0.2} & 1     & 0.420362 & -0.159482 & 0.4214 & -0.1579 & -0.246\% & 1.002\% \\
          & 2     & 0.776690& -0.164032 & 0.7763 & -0.1624 & 0.050\% & 1.005\% \\
    \hline
    \multirow{2}[0]{*}{0.5} & 1     & 0.368757 & -0.150491 & 0.3703 & -0.1492 & -0.417\% & 0.865\% \\
          & 2     & 0.688460& -0.155641 & 0.6876 & -0.1549 & 0.125\% & 0.478\% \\
    \hline
    \multirow{2}[0]{*}{1} & 1     & 0.367154 & -0.182131 & 0.3687 & -0.1805 & -0.419\% & 0.904\% \\
          & 2     & 0.710222 & -0.192072 & 0.7111 & -0.1899 & -0.123\% & 1.144\% \\
    \hline
    \multirow{2}[0]{*}{2} & 1     & 0.442449 & -0.371429 & 0.4529 & -0.3697 & -2.308\% & 0.468\% \\
          & 2     & 1.01997 & -0.436967 & 1.0278 & -0.4280 & -0.762\% & 2.095\% \\
    \hline
    \end{tabular}
    \caption{Error comparison between the WKB method and the Prony method for vector field perturbations. The relative error is calculated using the Prony method as the benchmark.}
    \label{Tab:4}
\end{table}

%\clearpage

%%%%%%%%%%%%%%%%%%%%%%%%%%%%%%%%%%%%%%%%%%%%%%%%%%%%%
\section{Comparison and discussion}
\label{sec:comparison}
%%%%%%%%%%%%%%%%%%%%%%%%%%%%%%%%%%%%%%%%%%%%%%%%%%%%%

In this section, we compare the QNMs of BHs in $f(\mathbb{Q})$ gravity with those in $f(\mathbb{T})$ gravity, highlighting the key distinguishing characteristics between the two frameworks. Furthermore, we propose a practical approach for utilizing real observational data to evaluate which of the three representative solutions (arising from these two theories) is more consistent with physical reality, thereby offering insights into which gravitational theory may be closer to the fundamental description of gravity.
The two gravity theories are expressed in the form of the series expansion and preserved to a quadratic term, 
that is, $f(\mathbb{Q})=\mathbb{Q}+\alpha \mathbb{Q}^2$ and 
$f(\mathbb{T})=\mathbb{T}+\beta \mathbb{T}^2$.
The reason that we ignore $f({R})$ gravity is 
$f({R})={R}+\gamma {R}^2$ 
only gives a correction term which is proportional to ${1}/{r}$ in the shape function, 
which can be treated as a correction of mass. 
Here $\beta$ and $\gamma$, similar to $\alpha$, are the deviation parameters showing the deviation from the linear theories.
Thus, $f({R})$ gravity cannot give any nontrivial result beyond GR, 
but just provides~\cite{Konoplya:2011qq} the effect of mass on QNMs. 
The following results on $f(\mathbb{T})$ gravity come from Ref.~\cite{Zhao:2022gxl}.

\begin{figure}[!ht]
     \centering
     \begin{subfigure}[b]{0.49\textwidth}
         \centering
         \includegraphics[width=\textwidth]{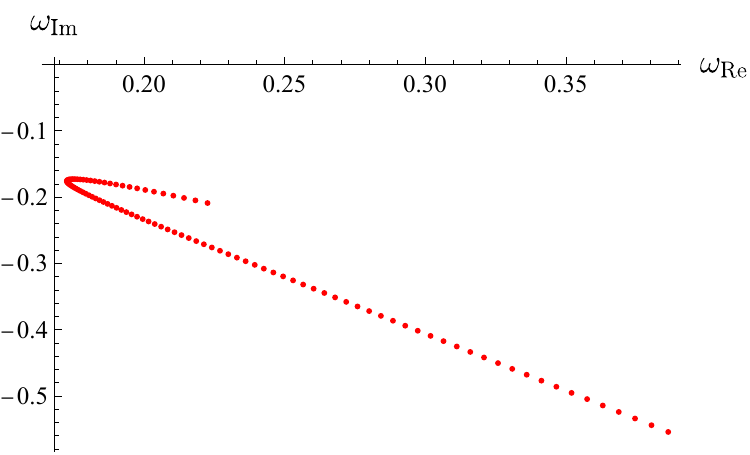}
         \caption{$f(\mathbb{Q})$ gravity.}
         %\label{fig:sca-per}
     \end{subfigure}
     %\hfill
     \begin{subfigure}[b]{0.49\textwidth}
         \centering
         \includegraphics[width=\textwidth]{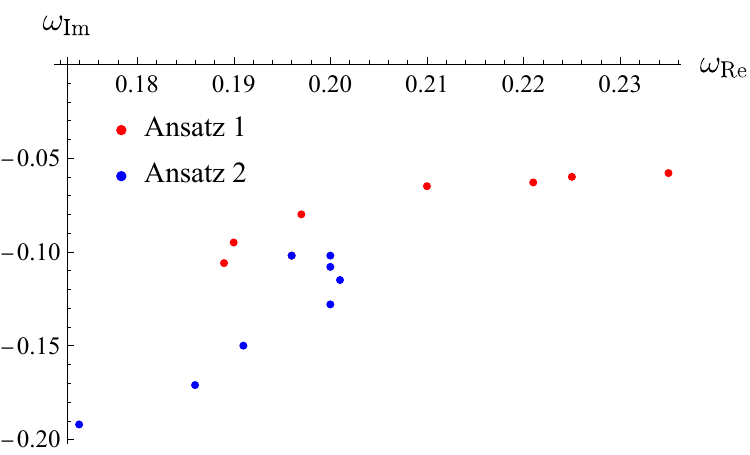}
         \caption{$f(\mathbb{T})$ gravity.}
         %\label{fig:ele-per}
     \end{subfigure}
      \captionsetup{width=.9\textwidth}
       \caption{The spectra of QNMs in both $f(\mathbb{Q})$ gravity and $f(\mathbb{T})$ gravity, where the parameters $\alpha$ and $\beta$ vary from $0$ to $2$, respectively, under a massless scalar field perturbation. Diagram (a) for $f(\mathbb{Q})$ gravity, where $l=0$, $n=0$, $k=1$, and $r_0=1$ are fixed, is based on the data in Table~\ref{Tab:1}. Diagram (b) is from Table 1 of Ref.~\cite{Zhao:2022gxl} for $f(\mathbb{T})$ gravity, where Ansatz 1 and Ansatz 2 are corresponding to the two families of solutions.}
        \label{fig:spec}
\end{figure}

The QNM spectra for both  $f(\mathbb{Q})$  gravity and  $f(\mathbb{T})$  gravity are presented in Figure ~\ref{fig:spec}. In  $f(\mathbb{Q}) $ gravity, $\omega_\mathrm{Re}$ exhibits a minimum with respect to $\omega_\mathrm{Im}$, and both $\omega_\mathrm{Re}$ and $\omega_\mathrm{Im}$ are double-valued relative to each other. When we compare the QNM spectrum of $f(\mathbb{Q})$  gravity with that of  $f(\mathbb{T})$  gravity, we find that  the spectrum of ansatz 1 in  $f(\mathbb{T})$ resembles that of  $f(\mathbb{Q})$  gravity, while the spectrum of ansatz $2$ in  $f(\mathbb{T})$ shows a maximum in $\omega_\mathrm{Re}$.

Notably, the oscillatory decay in the late-stage waveform of ansatz $1$ suggests that it may be possible to distinguish among the three solutions observationally. 
That is, we analyze whether the tail of ringdown waveforms is free of oscillations or not, as well as the opening direction of the QNM spectrum (as shown in Figure~\ref{fig:spec}). The solutions we consider have the following characteristics: $f(\mathbb{Q})$ gravity — tail without oscillations, QNM spectrum opens to the right; $f(\mathbb{T})$ gravity Ansatz 1 — tail with oscillations, QNM spectrum opens to the right; $f(\mathbb{T})$ gravity Ansatz 2 — tail without oscillations, QNM spectrum opens to the left. It is worthy mentioning that
the distinctive QNM spectrum features in  $f(\mathbb{Q})$  and  $f(\mathbb{T})$  gravity are reminiscent of patterns associated with phase transitions, such as those occurred in Reissner–Nordstr\"om black holes \cite{Jing:2008an}. Specifically, the turning point in Figure ~\ref{fig:spec} may correspond to the Davies point, which marks second-order phase transitions \cite{Lan:2020fmn}. Further investigation from a thermodynamic perspective is required to confirm this connection, which we plan to address in our future work.

In contrast to $f(\mathbb{Q})$ gravity, the QNMs of the two families of spherically symmetric BH solutions in $f(\mathbb{T})$ gravity appear less sensitive to the deviation parameter, though some correlations remain evident. As shown in Figure ~7 of Ref.~\cite{Zhao:2022gxl}, the absolute values of the real and imaginary parts of the QNMs in $f(\mathbb{T})$ gravity decrease as the deviation parameter increases, resulting in lower oscillation frequencies and slower decay rates.

It is obvious that, for the same $M$ and $l$, the sensitivity of QNMs to  the deviation parameter is significantly more pronounced in $f(\mathbb{Q})$ gravity, with the relationship between the real and imaginary parts of the QNMs exhibiting an opposite trend compared to $f(\mathbb{T})$ gravity. This distinction provides a potential observational means to differentiate between the two theories.

Furthermore, it is worth noting that the first family of solutions in Ref.~\cite{Zhao:2022gxl} slightly differs from the solutions in $f(\mathbb{Q})$ gravity, where the former is characterized by a non-zero effective potential near the horizon. In contrast, the second family’s effective potential approaches zero near the horizon, leading to a slow power-law decay tail similar to that observed in $f(\mathbb{Q})$ gravity.

\begin{figure}[!ht]
     \centering
         \includegraphics[width=.6\textwidth]{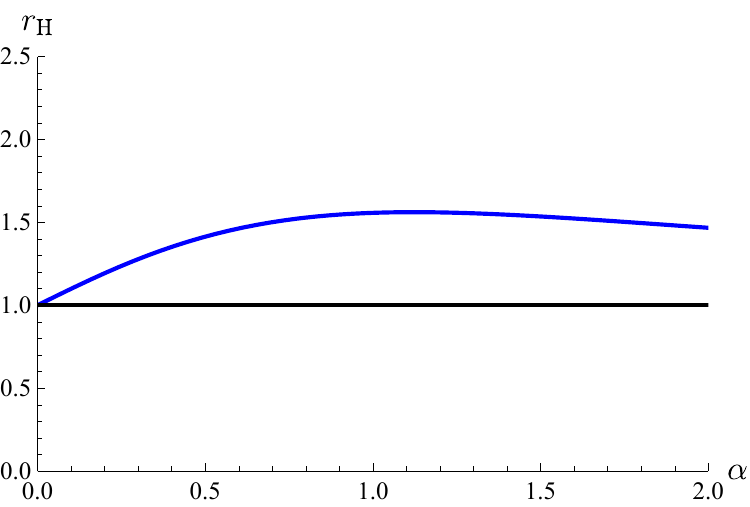}
      \captionsetup{width=.9\textwidth}
       \caption{The relationship between the horizon radius and $\alpha$ in $f(\mathbb{Q})$ gravity (blue curve), where $l=0$, $n=0$, $k=1$, and $r_0=1$ are fixed. Note that $r_0$ (black line) always remains inside the horizon ($r_{\rm H}\ge 1$).}
        \label{fig:horizon}
\end{figure}

For fixed $k=1$ and $r_0=1$, the parameter $r_0$ remains inside the event horizon, see Figure ~\ref{fig:horizon}. We attribute this behavior to the influence of the $\alpha^2$ term of Eq.~(\ref{eq:f(r)}), which plays a critical role in shaping the system’s dynamics.

Moreover, when the angular momentum quantum number $l$ increases, the real part of QNMs becomes increasingly sensitive to $\alpha$, while the imaginary part does conversely. This highlights the intricate dependence of QNMs on both $\alpha$ and $l$, emphasizing their significance in determining the dynamics of BHs in $f(\mathbb{Q})$ gravity.

\begin{remark}
Understanding physical roles of every parameters, is a necessary prerequisite for any possible future phenomenological study.
In this section, we aim to systematically discuss the potential physical interpretations of the three parameters appearing in our metric, Eq.~\eqref{eq:f(r)}. Although a detailed and rigorous analysis of their physical significance lies beyond the scope of the present work, we offer several plausible considerations and insights.

We begin with the parameter $\alpha$, which depicts the deviation from the linear theory of $f(\mathbb{Q})$ gravity.
We confirm that $\alpha$ is associated with the QNM spectrum, potentially establishing a connection between theoretical predictions and future observational tests.

Next, we consider the parameter $k$. From the structure of the metric in Eq.~\eqref{eq:f(r)},
we know that $k$ must have the dimension of inverse mass. Moreover, since $k$ is constrained to be positive and appears as the coefficient of the leading correction term in $f(\mathbb{Q})$ gravity, i.e., the logarithmic term, 
it effectively measures the strength or scale of modifications introduced by $f(\mathbb{Q})$ gravity.

Finally, the interpretation of the parameter $r_0$ is somewhat more subtle, whose presence is to ensure that the argument of the logarithmic term remains dimensionless. 
Furthermore, the fact that the sign of the logarithmic term depends on the relative magnitude of $r$ and $r_0$ suggests that $r_0$ may indicate a spatial change, particularly in the region outside the black hole. Given that the logarithmic correction becomes dominant at large distances, i.e., $\alpha < \alpha^2 k\ln(r)$, such effects may have observable consequences in the asymptotic regime.

\end{remark}

%%%%%%%%%%%%%%%%%%%%%%%%%%%%%%%%%%%%%%%%%%%%%%%%%%%%%
\section{Conclusion}
\label{sec:conclusion}

%%%%%%%%%%%%%%%%%%%%%%%%%%%%%%%%%%%%%%%%%%%%%%%%%%%%%

In this work, we analyze the QNMs of static and spherically symmetric BHs in $f(\mathbb{Q})$ gravity 
under a massless scalar and electromagnetic field perturbations.
We consider a quadratic correction of $\mathbb{Q}$, $f(\mathbb{Q}) = \mathbb{Q} + \alpha \mathbb{Q}^2$, 
which serves as a reasonable approximation for any series expansion of $f(\mathbb{Q})$. 
To simplify the free parameters, we fix $M = 1/2$, which means that the mass primarily acts as a scaling factor for the QNM frequencies $\omega$.

By using  the tortoise coordinate, 
we reformulate the general perturbation equations into a Schr\"odinger-like form, 
enabling QNM calculations via the finite difference and WKB methods. 
We extract the effective potentials governing wave evolution, 
which universally approach zero at infinity and 
reach a maximum around $r^* = 3/2$. 
Notably, regardless of parameter values, the effective potential near the horizon always vanishes. 
This behavior contrasts with that of $f(\mathbb{T})$ gravity \cite{Zhao:2022gxl}, 
where one solution yields a positive effective potential near the horizon, 
leading to oscillatory power-law decay at fixed frequencies in the wave tail, a clear observational distinction between $f(\mathbb{T})$ and $f(\mathbb{Q})$.

We calculate the QNMs with the finite difference method and the six-order WKB plus Pad\'e approximation, 
give the time evolution of perturbation waves, and analyze the relationships between QNM frequencies and the parameters, like the angular momentum $l$, 
the correction coefficient $\alpha$, the integral constant $k$, and $r_0$, respectively. In particular, we emphasize the physical significance of these parameters.
Our results show that $f(\mathbb{Q})$ gravity introduces QNM features distinct from those of GR, 
with a strong dependence on the correction coefficient and other parameters. 
A notable feature of QNMs in $f(\mathbb{Q})$ gravity is the ``closed” spectrum, 
where the real and imaginary parts are double-valued with respect to each other. 
This means that the different decay rates may be observed for the same oscillation frequency, 
and vice versa, which is an important observational signature of $f(\mathbb{Q})$ gravity.

Our comparisons to $f(\mathbb{T})$ gravity with a quadratic correction 
reveal that the QNMs in $f(\mathbb{Q})$ gravity are more sensitive to parameter changes. 
Moreover, the relationship between the QNMs and the correction coefficient in $f(\mathbb{Q})$ gravity is opposite to that in $f(\mathbb{T})$ gravity. These differences may offer a promising basis for distinguishing the two theories through observations. 
It is worth mentioning that the first family of solutions in $f(\mathbb{T})$ gravity has a different waveform from that in $f(\mathbb{Q})$ gravity, while the second family of solutions in $f(\mathbb{T})$ gravity has the different QNM spectrum from that in $f(\mathbb{Q})$ gravity. It offers a practical way to distinguish them under observation.

It is worthy to notice that we use the Klein-Golden equation in the Levi-Civita connection when we derive the perturbation equation. 
This is a widely accepted convention in the study of quasinormal modes of BHs under modified gravity \cite{Zhao:2022gxl,Gogoi:2023kjt,Al-Badawi:2024iqv}. 
Although the Levi-Civita connection is consistent with the coincident gauge in  our consideration,  
it is an approximation in general since the connection under a non-Riemannian geometry may differ from the Levi-Civita connection and may have a relationship with parameters in different models. 
This issue becomes particularly significant when the connection used to solve black hole solutions differs from the one employed for test-field perturbations. 
Such inconsistency can lead to discrepancies in the analysis. Therefore, it is a critical challenge to establish a self-consistent and unified connection for studying the QNMs of test-field perturbations.

Considering the nonzero effect of effective potentials  near horizons in $f(\mathbb{T})$, we
make a brief analysis of this feature in Appendix, where our purpose is to avoid fake nonzero behaviors of effective potentials near horizons in $f(\mathbb{Q})$ and other gravitational theories.

In summary, we provide a comprehensive analysis of QNMs in static and spherically symmetric BHs under the framework of $f(\mathbb{Q})$ gravity and highlight several key differences from $f(\mathbb{T})$ gravity theory. 
Our findings offer a theoretical foundation for evaluating the validity of modified gravity theories and 
contribute valuable insights into investigating QNMs in alternative theories of gravity.
Future work may focus on the QNM properties of rotating BHs in $f(\mathbb{Q})$ gravity.

%%%%%%%%%%%%%%%%%%%%%%%%%%%%%%%%%%%%%%%%%%%%%%%%%%%%%%%%%%%%
\section*{Acknowledgements}
%%%%%%%%%%%%%%%%%%%%%%%%%%%%%%%%%%%%%%%%%%%%%%%%%%%%%%%%%%%%

The authors would like to thank Hao Yang and Zhong-Wu Xia for their helpful discussions.
This work was supported in part by the National Natural Science Foundation of China under Grant No.\ 12175108. L.C. is also supported by Yantai University under Grant No.\ WL22B224.
Z.-X. Z is also supported by the Pilot Scheme of Talent Training in Basic Sciences (Boling Class of Physics, Nankai University), Ministry of Education.

%%%%%%%%%%%%%%%%%%%%%%%%%%%%%%%%%%%%%%%%%%%%%%%%%%%%%
\section*{Appendix: Discussion on the nonzero effective potential near horizons}
%%%%%%%%%%%%%%%%%%%%%%%%%%%%%%%%%%%%%%%%%%%%%%%%%%%%%

At the end of Sec.\ \ref{sec:sssol}, we mention the nonzero behavior of the effective potential near the horizon in $f(\mathbb{T})$ gravity, as discussed in Ref.\ \cite{Zhao:2022gxl}. It is attributed to the non-equivalence of roots of $g_{00}$ and $1/g_{11}$. 
Such nonzero effective potentials near the horizon commonly occur in spherically symmetric black holes with the algebraic property $[1, 1(11)]$,\footnote{This refers to the Segr\'e classification of spacetime. For a detailed explanation, see Appendix A of Ref.~\cite{Lan:2020fmn}.} where the metric components $g_{00}$ and $1/g_{11}$ exhibit distinct roots.

\begin{remark}
    
    We point out that a fake nonzero behavior of the effective potential in $f(\mathbb{Q})$ gravity may occur when one takes an unsuitable approximation of $r^*$. For example, if we take
    \begin{equation}
        r^* = r+\ln (r-1) +\alpha  \left[\frac{1}{1-r}+\ln (r-1)\right]+ \mathcal{O}(\alpha^2),
    \end{equation}
    as the approximation of $r^*$, instead of our correct choice Eq.\ \eqref{approxtorcor}, this approximation gives rise to two possible deformations of the effective potentials, as shown in Figure \ \ref{fig:deform}.
    
    \begin{figure}[!ht]
	\centering
	\includegraphics[width=0.6\linewidth]{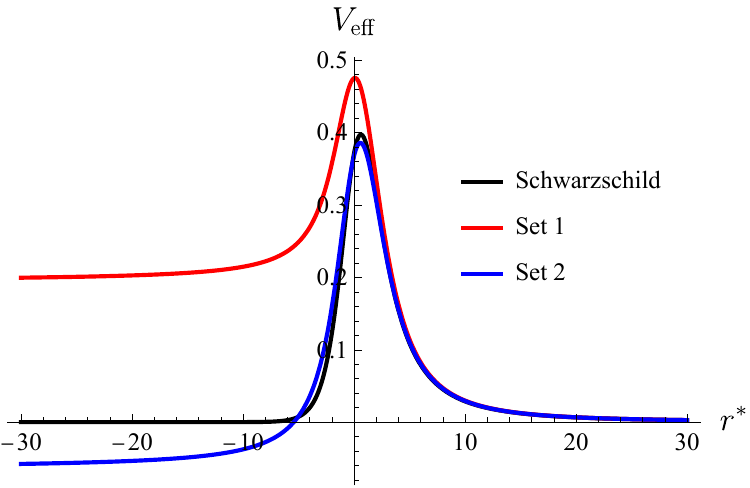}
	\captionsetup{width=.9\textwidth}
	\caption{Deformations of the effective potential for the scalar field perturbation under two sets of parameters in an unsuitable approximation: Set 1 ($\alpha = 0.05$,  $k = 20$,  $r_0 = 0.1$) and Set 2 ($\alpha = 0.05$,  $k = 5$,  $r_0 = 0.1$).}
	\label{fig:deform}
    \end{figure}

    Numerical calculation errors and improper truncation may also lead to fake nonzero behavior of the effective potential, thereby affecting the analysis of ringdown waveform characteristics. We recommend conducting universal checks in numerical calculations and comparing them with the analytical form of the effective potential when processing similar issues under any gravitational theories.
    
\end{remark}

\begin{figure}[!ht]
	\centering
	\includegraphics[width=.6\textwidth]{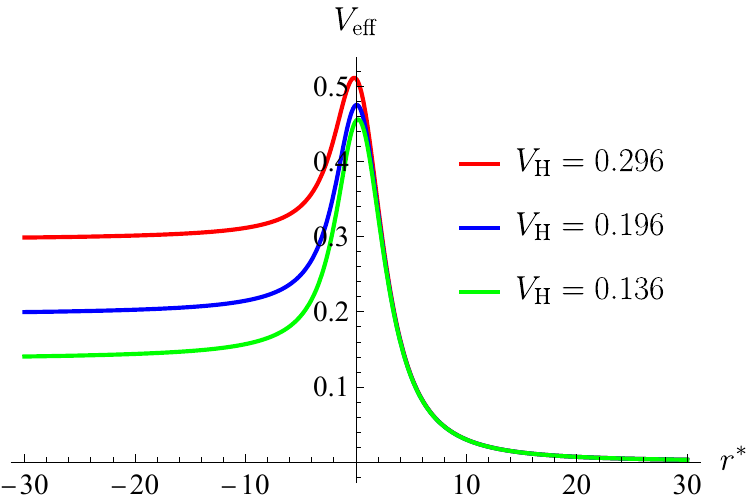}
	\captionsetup{width=.9\textwidth}
	\caption{Effective potential with respect to the tortoise coordinate, where the asymptotic values near horizons, $V_{\rm H}$, are positive.}
	\label{fig:altails_Veff}
\end{figure}

\begin{figure}[!ht]
     \centering
         \includegraphics[width=.6\textwidth]{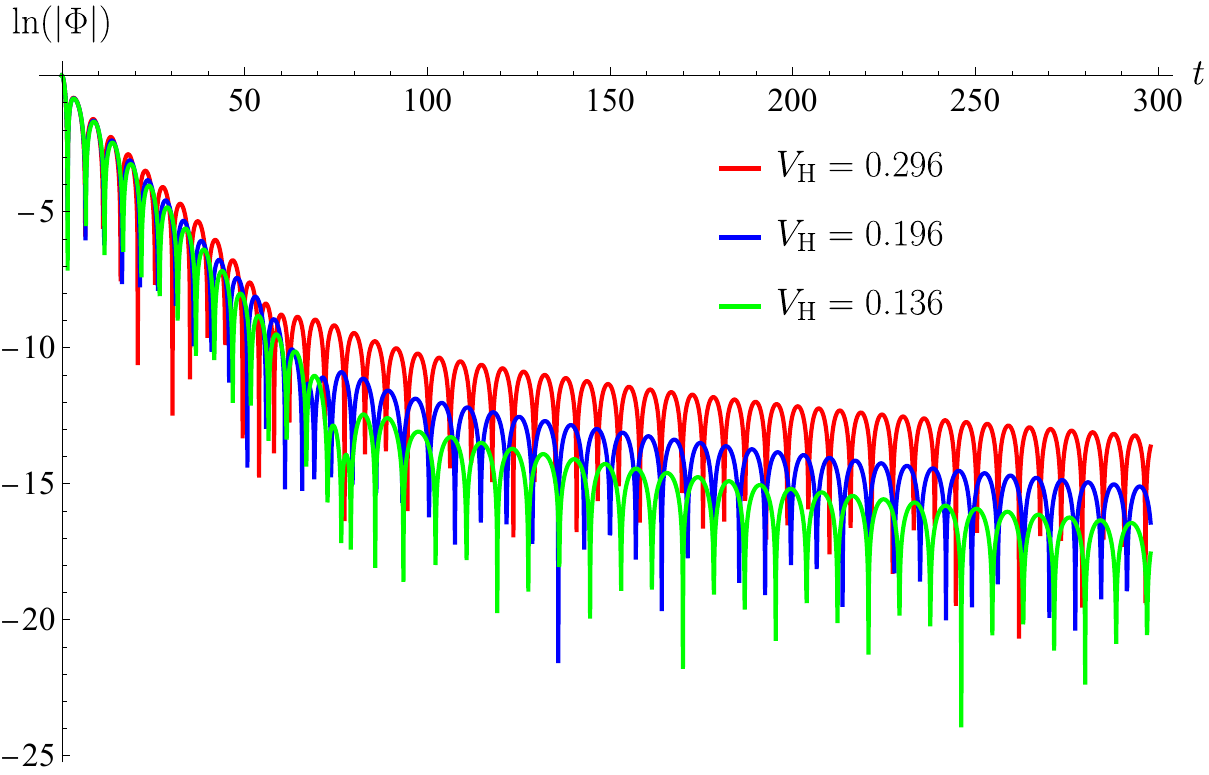}
      \captionsetup{width=.9\textwidth}
       \caption{Waveform of scalar field perturbations for positive asymptotic values of effective potentials near horizons.}
        \label{fig:altails}
\end{figure}

In the case that the effective potential near the horizon ($V_{\rm H}$) is positive, see  Figure ~\ref{fig:altails_Veff}, 
the waveform tails display a unique oscillatory decay, see Figure \ \ref{fig:altails}, which differs from the purely decaying tails shown in Figs.\ 
\ref{fig:diff_rc} %\ref{fig:diff_rTorAssigned} 
and \ref{fig:diff_parameters}. In particular,  Figure ~\ref{fig:altails} shows that a higher $V_\mathrm{H}$ causes a higher intensity and a greater frequency in the oscillatory decay of the tail period. 

\begin{figure}[!ht]
	\centering
	\includegraphics[width=.6\textwidth]{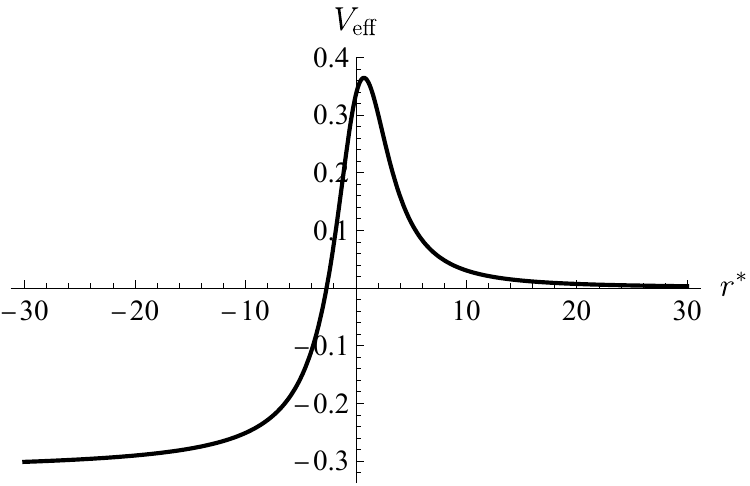}
	\captionsetup{width=.9\textwidth}
	\caption{Effective potential which has a negative asymptotic value at the horizon. Here we fix $\alpha=0.1$, $k=10$, $r_0=1$, and $l=1$.}
	\label{fig:nega_Veff}
\end{figure}

\begin{figure}[!ht]
     \centering
         \includegraphics[width=.6\textwidth]{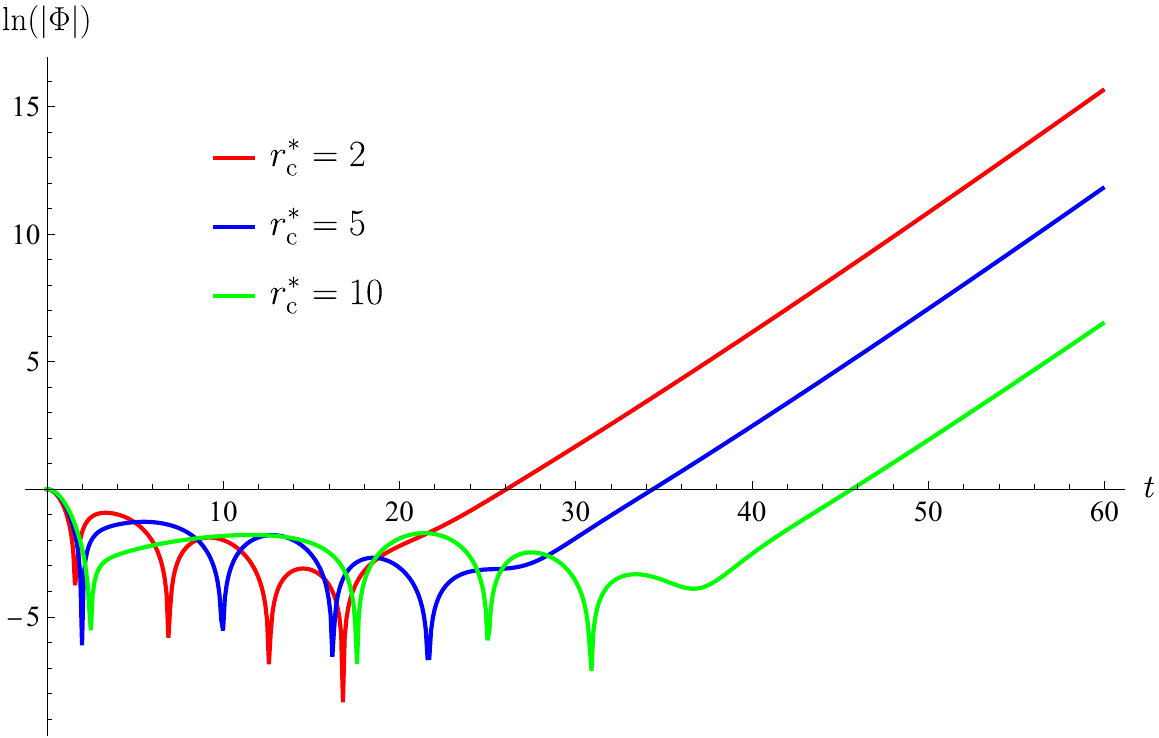}
      \captionsetup{width=.9\textwidth}
       \caption{Waveform of perturbation fields for negative asymptotic values of effective potentials near horizons, where the tails grow  exponentially.}
        \label{fig:nega_time_domain}
\end{figure}

In the case that the effective potential near the horizon is negative, the approach used for the positive case leads to a divergent waveform tail. For instance, when we set $\alpha = 0.1$, $k = 10$, $r_0 = 1$, and $l = 1$, the effective potential exhibits the behavior shown in Figure \ \ref{fig:nega_Veff}, where the potential approaches a negative value as $r^* \to -\infty$. We then apply the finite difference method to analyze the time-domain behavior of the perturbation equation, and give the results displayed in Figure \ \ref{fig:nega_time_domain}. That is, after an initial delay, positively correlated with the distance from the horizon of black holes, the perturbation fields grow exponentially. This behavior may be compared with that of the Schr\"odinger-like solution, $\Phi(r, t) = \phi(r) \exp\left(-{\mi}E t/{\hbar} \right)$, where $E$ is complex. If the imaginary part of $E$ is positive (corresponding to a negative potential), the evolution of time turns out to be divergent, leading to an unstable state. The similar phenomenon was also seen in the BHs of Einstein-Gauss-Bonnet theory \cite{cuyubamba2016quasinormal,cuyubamba2021stability}.

The above divergent tail is clearly unacceptable because it means that the perturbation field carries an infinite energy. Here we propose three possible interpretations to address it.
The first is that any parameter choice resulting in a negative effective potential is inherently nonphysical or invalid. 
It provides a straightforward resolution and imposes stricter constraints on parameter selections in modified gravity theories.
The second interpretation is to accept the phenomenon by considering that the test field could draw energy from a black hole, 
leading to a reduction of black hole’s ADM mass. 
In some cases, this process might allow the black hole changes to a stable phase, 
after which the test field would begin to decay again. 
Under this view, such parameter choices would represent unstable but not entirely prohibited configurations. 
However, for other parameter choices, where the black hole’s mass could decrease indefinitely (potentially exposing a naked singularity), 
these configurations would still be deemed unphysical.
The third interpretation suggests that the effective potential should instead be considered in terms of its absolute value. 
While it offers a different perspective, 
its validity requires further investigation.
The most appropriate resolution for this phenomenon remains an open question.
Since this issue lies beyond the primary scope of this work, 
we defer its detailed discussion to future studies.

\bibliographystyle{utphys}
\bibliography{references}
\end{document}